\documentclass[fleqn,usenatbib]{mnras}
\usepackage{newtxtext,newtxmath}
\usepackage[T1]{fontenc}
\usepackage{xspace}

\usepackage{graphicx}	%
\usepackage{amsmath}	%
\usepackage[]{siunitx}
\usepackage[export]{adjustbox}
\usepackage{fixme}
\fxsetup{
  status=draft,
  author=CC,
  layout=inline,nomargin,nofootnote,
  theme=colorsig,
}
\usepackage{enumitem}

\FXRegisterAuthor{cc}{acc}{CC}
\FXRegisterAuthor{mr}{amr}{MR}	%
\newcommand{\mynotes}[1]{}
\newcommand{\TODOlater}[1]{}              %

\newcommand{\hide}[1]{}         %
\newcommand{\edit}[1]{{#1}}

\newcommand{\ie}{\textit{i.e.}}
\newcommand{\eg}{\textit{e.g.}}
\newcommand{\ramses}{\textsc{ramses-rt}}

\newcommand{\msun}{\ensuremath{{\rm M}_{\odot}}}
\newcommand{\pcc}{\ensuremath{{\rm cm}^{-3}}}	%

\newcommand{\cHH}{{\tt $\mu$3Ma15}}
\newcommand{\cMH}{{\tt $\mu$1Ma15}}
\newcommand{\cLH}{{\tt $\mu$0.6Ma15}}
\newcommand{\cHL}{{\tt $\mu$3Ma7}}
\newcommand{\cHR}{{\tt $\mu$3Ma7-hires}}  %
\newcommand{\cML}{{\tt $\mu$1Ma7}}
\newcommand{\cLL}{{\tt $\mu$0.6Ma7}}

\title[Formation of Large Discs in Magnetised Cores]{Formation of Large Circumstellar Discs in Multi-scale, ideal-MHD Simulations of Magnetically Critical Pre-stellar Cores} 

\author[C.-C. He, M. Ricotti]{
  Chong-Chong He,$^{1}$\thanks{E-mail: che1234@umd.edu}, Massimo Ricotti$^1$\thanks{E-mail: ricotti@umd.edu},\\
  $^{1}$Department of Astronomy, University of Maryland, College Park, MD, 20742, US\\
}
\date{Accepted XXX. Received YYY; in original form ZZZ}
\pubyear{2023}

\begin{document}
\label{firstpage}
\pagerange{\pageref{firstpage}--\pageref{lastpage}}
\maketitle

\begin{abstract}
The formation of circumstellar discs is a critical step in the formation of stars and planets. Magnetic fields can strongly affect the evolution of angular momentum during prestellar core collapse, potentially leading to the failure of protostellar disc formation. This phenomenon, known as the magnetic braking catastrophe, has been observed in ideal-MHD simulations. In this work, we present results from ideal-MHD simulations of circumstellar disc formation from realistic initial conditions of strongly magnetised, massive cores with masses between $30 ~{\rm M}_\odot$ and $300 ~{\rm M}_\odot$ resolved by zooming into Giant Molecular Clouds with masses $\sim 10^4 \ {\rm M}_\odot$ and initial mass-to-flux ratios $0.6 \le \mu_0 \le 3$. Due to the large turbulence caused by the non-axisymmetric gravitational collapse of the gas, the dominant vertical support of discs is turbulent motion, while magnetic and turbulent pressures contribute equally in the outer toroid. The magnetic field topology is extremely turbulent and incoherent, reducing the effect of magnetic braking by roughly one order of magnitude and leading to the formation of large Keplerian discs even in magnetically critical or near-critical cores. Only cores in GMCs with $\mu_0 < 1$ fail to form discs. Instead, they collapse into a sheet-like structure and produce numerous low-mass stars. We also discuss a universal $B-\rho$ relation valid over a large range of scales from the GMC to massive cores, irrespective of the GMC magnetisation. This study differs from the vast literature on this topic which typically focus on smaller mass discs with idealised initial and boundary conditions, therefore providing insights into the initial conditions of massive prestellar core collapse and disc formation.
\end{abstract}


\section{Introduction}

The formation of protostellar discs is ubiquitous during the collapse of prestellar cores. As a molecular core collapses under its own self-gravity, the angular momentum of the gas will slow down its collapse at small scales promoting the formation of a protostellar disc. This simple argument of conservation of angular momentum is corroborated by observations, also suggesting that protostellar disc formation is a natural byproduct of the star formation process \citep{Odell1992,Tobin2012,Murillo2013,Codella2014,Lee2017}. 

However, molecular clouds are observed to be permeated by magnetic fields \citep{Crutcher1999, Lee2017}, which can in principle strongly affect the evolution of angular momentum during the core collapse. The twisting of the magnetic field lines produced by the disc rotation in the flux-freezing regime of ideal magnetohydrodynamics (MHD), can apply a force counter to the rotation velocity, also known as magnetic braking, effectively slowing down rotation and increasing radial gas infall. 
In idealized numerical MHD simulations, the timescale of the braking can become so short that protostellar discs fail to form or are much smaller than the observed sizes, a phenomenon known as ``the magnetic braking catastrophe'' \citep[e.g.][]{Allen2003, Galli2006, Hennebelle2008a, Li2014}. Indeed, disc formation should be completely suppressed in the strict ideal MHD limit for the level of core magnetization deduced from observation -- the angular momentum of the idealised collapsing core is nearly completely removed by magnetic braking close to the central object \citep[\eg,][]{Mestel1956, Mellon2008}.
These results seem to be in contrast to the observed existence of Keplerian discs around protostellar objects. \edit{The observed sizes of Class 0 and Class I protostellar discs revealed by recent radio/mm and optical/IR observations \citep{Zapata2007,Sanchez-Monge2010,vanKempen2012,Takahashi2012,Johnston2015,Johnston2020}, have radii $\sim 100$ AU (and in some cases  $R \gtrsim 500$~AU) that are difficult to reconcile with ideal magneto-hydrodynamic models of disc formation, at least assuming idealized initial conditions \citep{Lebreuilly2021}.} 

Magnetic fields support charged gas against gravitational collapse. A common characterisation of the relative importance of the gravitational and magnetic forces in a molecular cloud or core is the normalised mass-to-flux ratio,
\begin{equation}\label{eq:mu}
  \mu \equiv \frac{M/\Phi_B}{M_{\rm \Phi} / \Phi_B}=\frac{M}{M_{\rm \Phi}},
\end{equation}
where $M$ is the total mass contained within a spherical region of radius $R$, $\Phi_B = \pi R^2 B$ is the magnetic flux threading the surface of the sphere assuming a uniform magnetic field strength $B$, and 
\begin{equation}\label{eq:mphi}
  M_{\rm \Phi} = c_{\Phi} \frac{\Phi_B}{\sqrt{G}},
\end{equation}
is the magnetic critical mass, the mass at which the magnetic and gravitational forces balance each other. The constant $c_{\Phi}$ is a dimensionless coefficient that depends on the assumed geometry of the system. For a spherical cloud of uniform density, $c_{\Phi} = \sqrt{10}/(6\pi) = 0.168$. However, the mass-to-flux ratio $\mu$ should be used with caution, since the definition of the critical value depends on the geometry of the gas and the magnetic fields.

In a sub-critical cloud (defined as a cloud with $\mu<1$), the magnetic field should prevent the collapse of the cloud core altogether. Observations suggest typical values of $\mu \approx 2 - 10$ in molecular cloud cores \citep[e.g.][]{Crutcher1999, Bourke2001}, and this value could be even smaller after correcting for projection effects \citep{Li2013}. Moreover, analytical predictions \citep{Joos2012} suggest that there are no centrifugally-supported discs in models with $\mu \le 10$, although there are pseudo-discs \edit{formed by infalling gas dragging and pinching the B-field lines hence producing a strong magnetic force counter-acting gravity \citep{Galli1993}.}

\edit{Simulating the formation and evolution of individual stars requires resolving the ``opacity limit'' for fragmentation -- the critical density above which the gas becomes optically thick to cooling radiation, leading to gas heating and a halt to fragmentation \citep{Rees1976}. This critical density is $10^{10} - 10^{11} \ \pcc{}$ at solar metallicities.
There is a vast literature studying collapsing gas near this density regime using both ideal MHD simulations \citep{Santos-Lima2012, Joos2012,Li2013c, Joos2013, Seifried2013, Li2014, Fielding2015, Gray2018, Tsukamoto2018, Lewis2018, Lam2019, Hennebelle2009, Wurster2020, Hirano2020}, or non-ideal MHD studies/simulations \citep{Machida2011,Masson2016,Wurster2016a,Kolligan2018,Zhao2018,Wurster2019, Tsukamoto2020,Kuffmeier2020,Mignon-Risse2021,Commercon2022,Mignon-Risse2023} . 
We refer to \cite{Tsukamoto2023} for a comprehensive review on this topic. 
Each of these previous studies, in addition to facing specific constraints due to computational cost/limitations, focus on different physics questions and has initial conditions that typically apply to describing traditional cores/circumstellar discs around solar type stars. Typically, the mass of the molecular clouds studied does not exceed 500 $\msun{}$, with the most common limit of around 100 $\msun{}$. Moreover, most of these simulations usually assume idealised initial and boundary conditions for the discs, along with artificially introduced turbulence, which may not fully capture the complexities of the star formation processes in a realistic molecular cloud environment.}

\edit{The main incremental improvement of this work, when compared to existing literature, is the adoption of realistic initial and boundary conditions for the collapse of high-mass prestellar cores.}
Here we stress two points: i) the masses of the cores presented in this study of circumstellar disc formation are significantly larger than in most previous studies: between 30~M$_\odot$ to 300~M$_\odot$; 2) We simulate these core adopting realistic initial conditions by zooming into strongly magnetised cores within a coarser resolution simulations of a $\sim 10^4 \ \msun{}$ GMCs, along with realistic boundary conditions from the co-evolving GMC. This paper is an extension of the work presented in \cite{He2022} (hereafter, Paper~I), in which we studied the formation and fragmentation of high-mass prestellar cores and the formation of large circumstellar discs. In this work, we focus on the magnetic phenomenon and the effects of of varying the cloud magnetisation, in the ideal-MHD regime. In section~\ref{sec:disc} we discuss how our results compare to the extensive body of literature on ideal- and non-ideal MHD simulations of circumstellar discs.

The rest of this article is organised as follows. We describe the methods in our simulations in \S~\ref{sec:method}. We present the results in \S~\ref{sec:res}, including an analysis of why the magnetic braking catastrophe does not prevent disc formation. A discussion in light of previous literature is presented in \S~\ref{sec:disc}, and a summary and conclusions in \S~\ref{sec:sum}.

\section{Method} 
\label{sec:method}

\begin{table*}
  \centering 
  \begin{tabular}{lllccccccl}
    GMC/core name  &\edit{$m_{3000} (\msun{})$}&\edit{$m_{\rm 0.1pc}$ ($\msun{}$)}& {$\mu_{0}$} & {${\cal M}$} & B ($\mu$G) & {${\cal M}_{A}$} & {$\beta$} & {$\Delta x_{\rm min}$} {(AU)}  &\edit{$n_{\rm sink}$ (cm$^{-3}$)}\\
	\hline
	$\mu$3Ma15            & 30&12& 3   & 15 & 10 & 5 & 0.1 & 60  &$8 \times 10^8$\\
        $\mu$3Ma15-large$^*$  &130&30& 3   & 15 & 10 & 5 & 0.1 & 60  &$8 \times 10^8$\\	
        $\mu$1Ma15            & 40&30& 1   & 15 & 30 & 1.7 & 0.01 & 60   &$8 \times 10^8$\\
	$\mu$0.6Ma15          & 300&25& 0.6 & 15 & 50 & 1 & 0.004 & 60   &$8 \times 10^8$\\
	$\mu$3Ma7             & 27&10& 3   & 7  & 5  & 5 & 0.5 & 29   &$4 \times 10^9$\\
	$\mu$3Ma7-hires       &27&10& 3   & 7  & 5  & 5 & 0.5 & 7  &$6 \times 10^{10}$\\
        $\mu$1Ma7             & 100&15& 1   & 7  & 15 &  1.7 & 0.06 & 29   &$4 \times 10^9$\\
	$\mu$0.6Ma7           & 60&10& 0.6 & 7  & 25 & 1 & 0.02 & 29  
   &$4 \times 10^9$\end{tabular}
  \caption{List of zoom-in simulations presented in this paper. Col. (2): core mass based on gas above a density threshold of $3000 \ \pcc{}$. Col. (3): core mass based on gas within a radius of 0.1 pc from the density peak. All the cores are chosen as the first star-forming core in the corresponding GMC, except for the one with a $^*$, which is chosen from the later stage of the GMC evolution and it is a very massive core with over 100 \msun{}. }
  \label{tab:1}
\end{table*}

In Paper~I we have conducted a set of ``zoom-in'' radiation-MHD simulations of prestellar core formation and evolution within collapsing GMCs. In this work, we extend the set of simulations to explore the effects of stronger magnetic fields. We present a suite of six ``zoom-in'' simulations of prestellar cores in molecular clouds with varying magnetisation and turbulence. 
We summarise the key parameters of the simulated GMCs and cores in Table~\ref{tab:1}. Two GMCs, \cHH{} and \cHL{} are fiducial runs with the same initial magnetisation ($\mu_0 = 3.0$) as those presented in Paper~I and in \cite{He2019}, in which we present the results for the larger scale GMCs evolution. The two clouds have initial turbulent Mach numbers ${\cal M} = 15$ and ${\cal M} = 7$ and masses of $3 \times 10^4 \ \msun{}$ and $3 \times 10^3 \ \msun{}$, respectively. These two GMCs are the XS-C and M-C clouds in \cite{He2019}, respectively. \edit{They represent clouds with mass, density, and magnetisation similar to Milky Way GMCs.} They both have an initial average number density of \edit{200} \pcc{}. 
The other four GMCs on which we zoom in (\cMH{}, \cML{}, \cLH{}, and \cLL{}) have the same properties as the two fiducial clouds but increased magnetic field intensities with $\mu_0 = 1.0$ and $\mu_0 = 0.6$, as shown by the name of the run. 
We also list the Alfven Mach number ${\cal M}_A$ and the Plasma beta $\beta$ of the initial GMC, as well as the spatial resolution $\Delta x_{\rm min}$ of the zoom-in simulations. 

For all the six zoom-in simulations mentioned above, we zoom into the first cores forming in each GMC and study their subsequent collapse. In this work we also include two extra cores selected from Paper~I: 1) {\tt $\mu$3Ma7-hires} -- which corresponds to Core {\it A-hr} in Paper~I -- is a higher-resolution zoom-in version of \cHL{}; 2) {\tt $\mu$3Ma15-large} corresponds to Core {\it B} in Paper~I, which is a very massive core that forms in the later stage of star formation in the fiducial GMC \cHH{}. 
\begin{figure*}
  \newcommand\gap{\hspace*{-0.35in}}
  \newcommand\ww{1.9in}
  \newcommand{\textbox}[1]{\framebox[1.6in]{#1}}
  \setlength{\fboxrule}{0pt}
  \raggedright
  \framebox[0.2in]{} \textbox{$\mu$3Ma15}\textbox{$\mu$1Ma15}\textbox{$\mu$1Ma7}\textbox{$\mu$0.6Ma15}\\
  \centering
  \includegraphics[width=\ww]{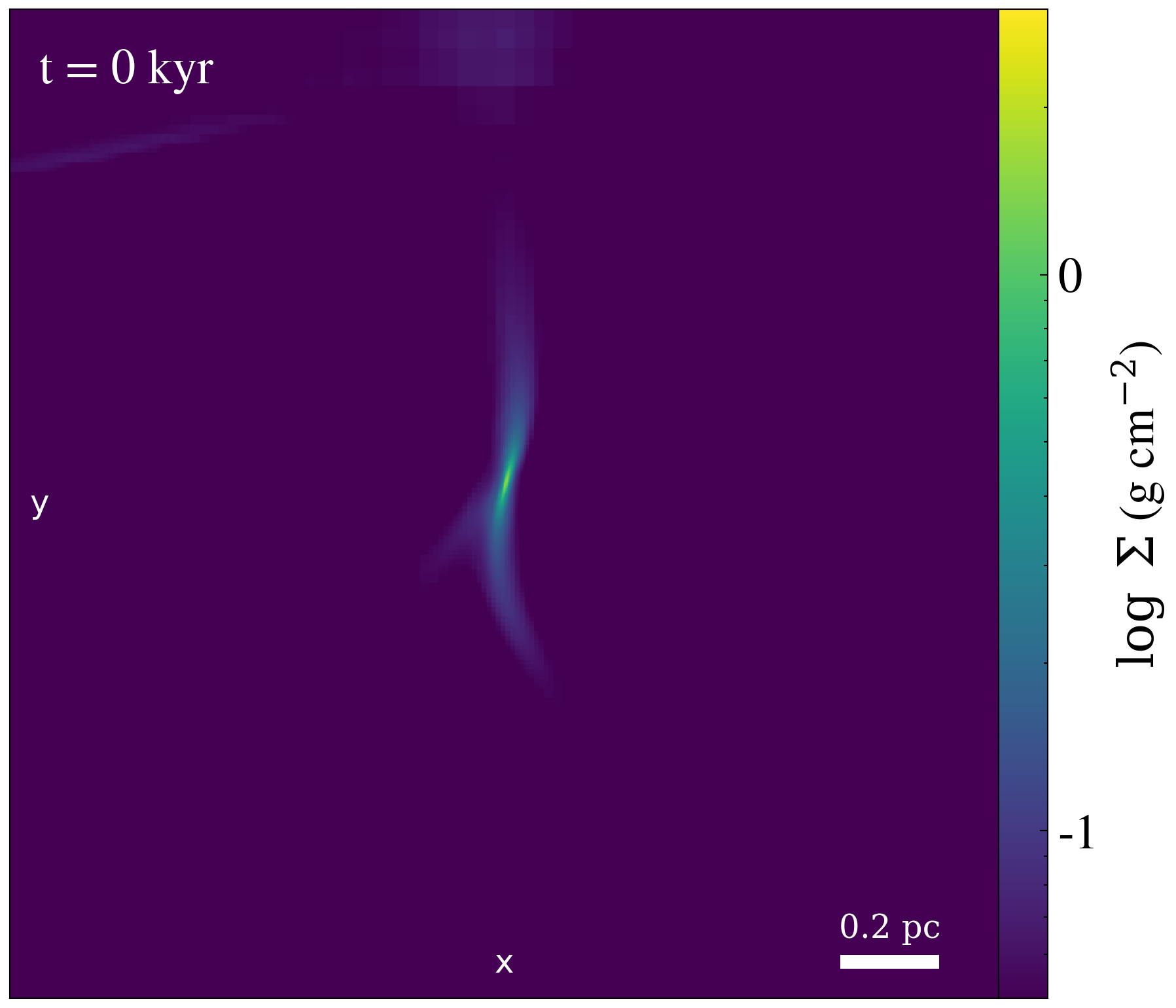}
  \gap{}
  \includegraphics[width=\ww]{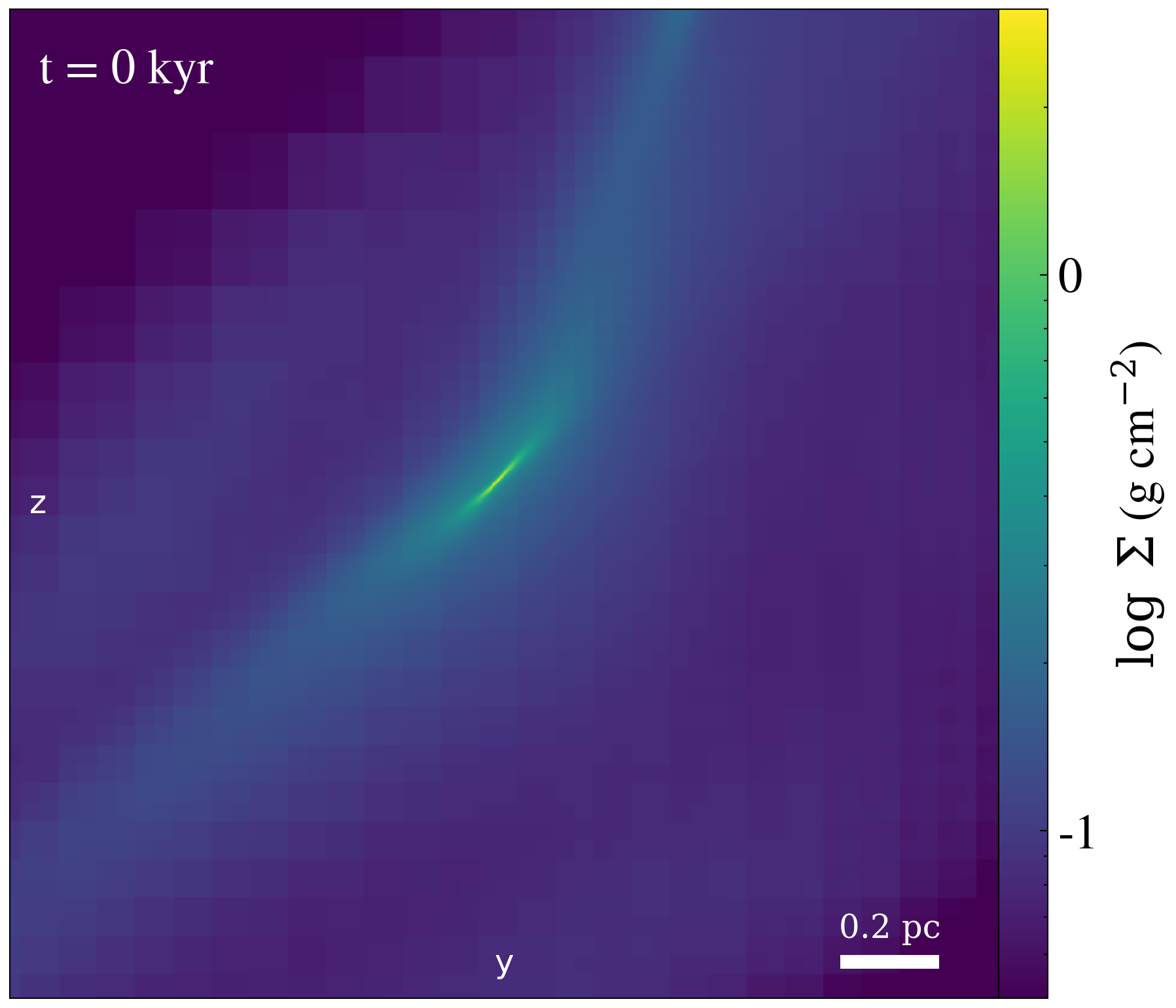}
  \gap{}
  \includegraphics[width=\ww]{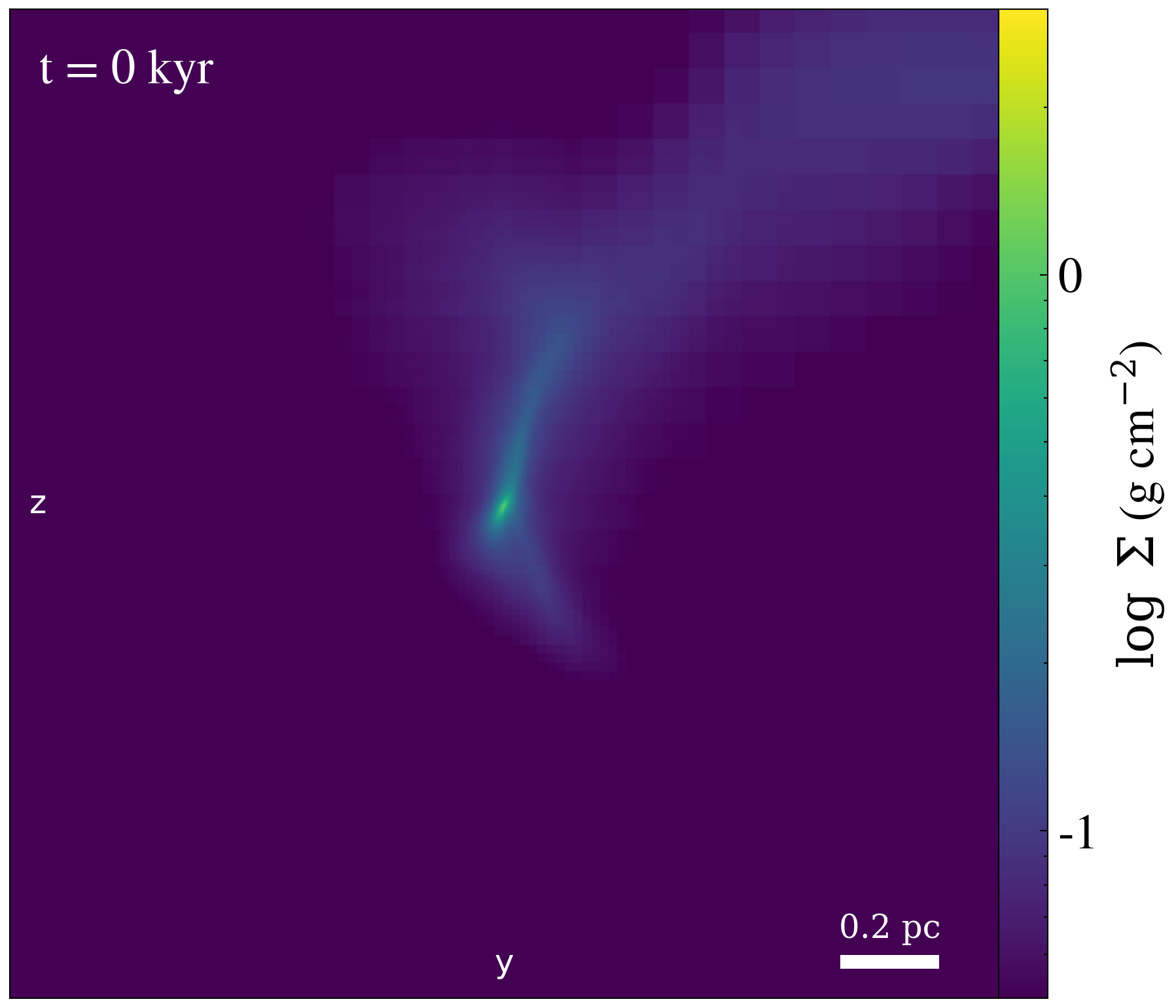}
  \gap{}
  \includegraphics[width=\ww]{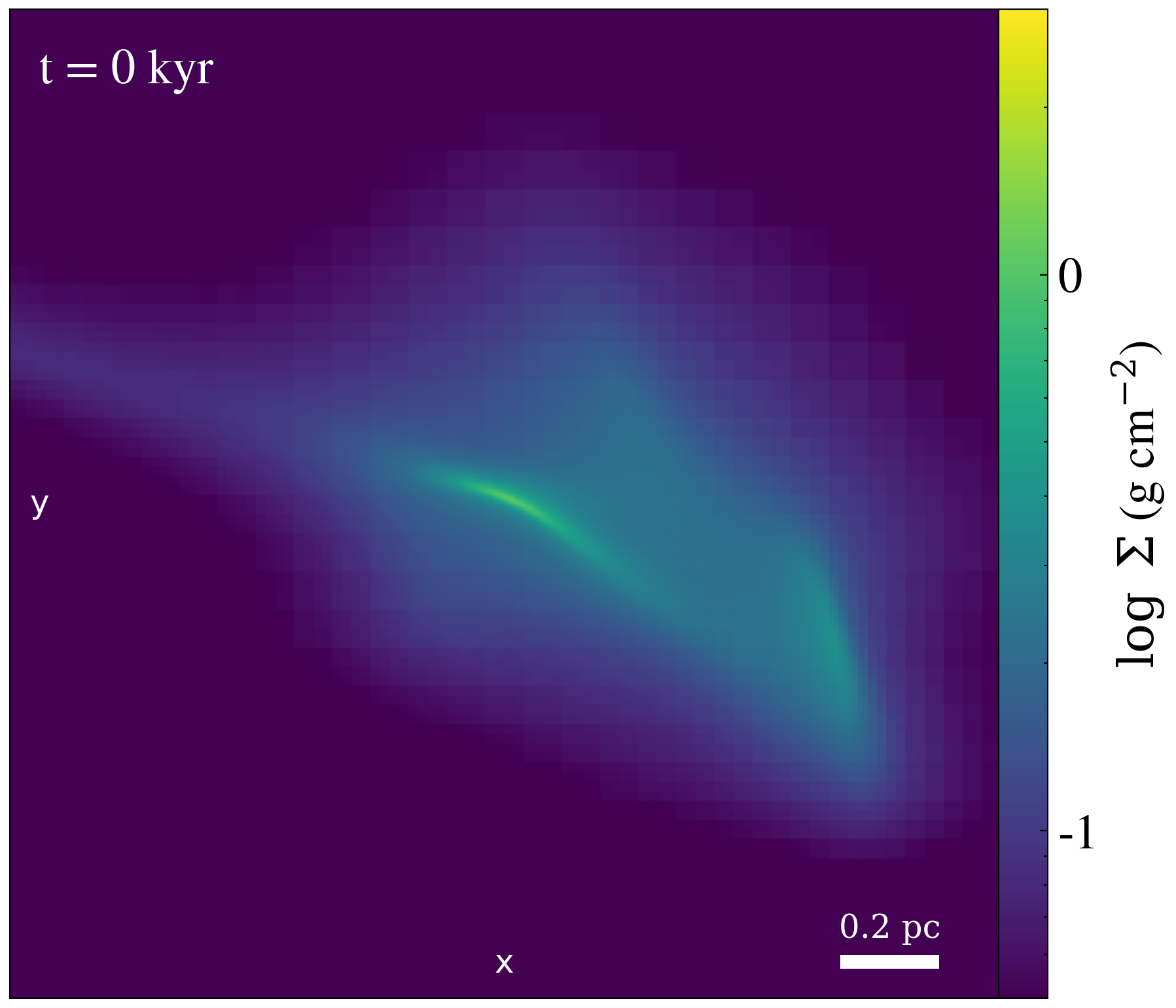}\\
  \includegraphics[width=\ww]{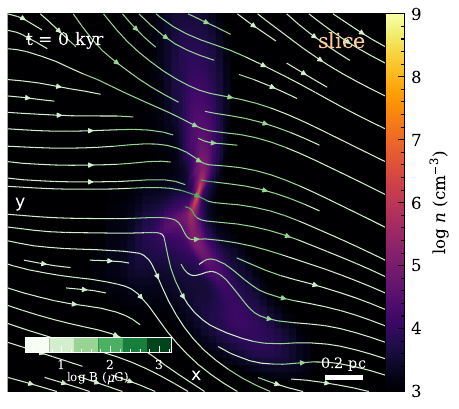}
  \gap{} 
  \includegraphics[width=\ww]{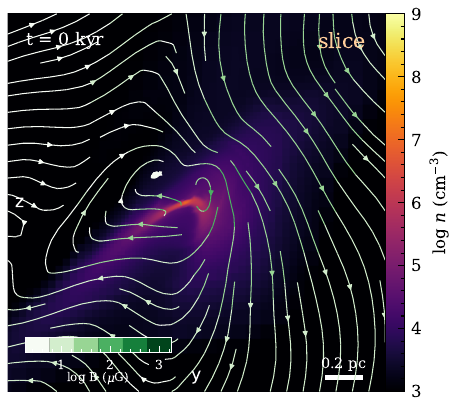}
  \gap{} 
  \includegraphics[width=\ww]{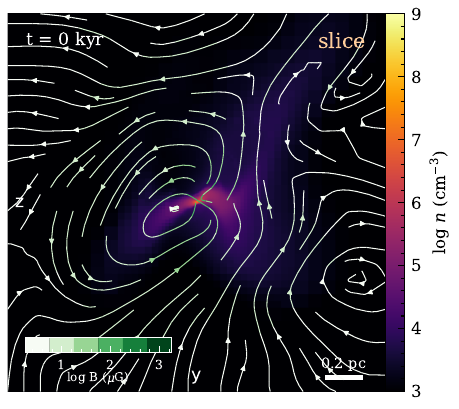}
  \gap{} 
  \includegraphics[width=\ww]{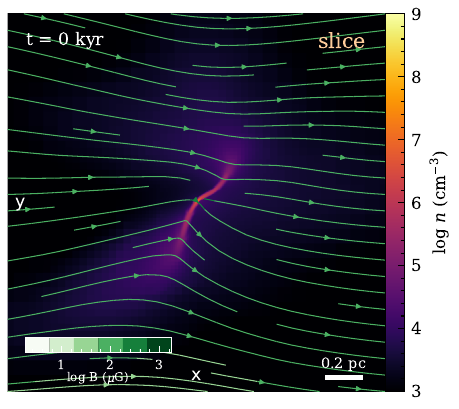}
  \caption{\label{fig:1} ({\it Top}) Surface density of the cores \cHH{}, \cMH{}, \cML{}, and \cLH{}, showing their morphology at $t = 0$, shortly before the formation of stars. ({\it Bottom}) Density slices of the corresponding cores centred at the peak density overplotted with magnetic field streamlines. The field of view and viewing angle orientation are the same as in the top panels. The magnetic field streamlines are colour-coded according to their magnitudes and a colorbar is shown at the bottom-left corner. Note how the surface density of the core \cLH{} does not intuitively reflect the actual geometry of its sheet-like shape.
  }
\end{figure*}

The sink particle (star formation) and stellar feedback recipes and the ``zoom-in'' method used in this work are described in Paper~I. Here, we provide only a brief summary. 
We perform simulations of star formation using the grid-based adaptive mesh refinement (AMR) MHD code \ramses{} \citep{Teyssier2002, Fromang2006}. Radiation transfer is modelled using a moment-based method with the M1 closure relationship for the Eddington tensor \citep{Rosdahl2013}. The ionising photons emitted from stars interact with neutral gas and we keep track of the time-dependent ionisation chemistry of atomic hydrogen and helium, but we do not include the chemical evolution of the molecular phase and metal chemistry, used only for cooling/heating rates, is treated assuming equilibrium abundances. Heating from photoionization and cooling from hydrogen and helium, metals, and dust grains are implemented. Cooling below 10~K is shut down to keep the temperature ﬂoor at 10 K. The photoionization feedback from stars heats the gas, dispersing the cloud and quenching star formation.

The baseline simulations of GMCs are started from idealised spherical isothermal clouds in hydrostatic equilibrium surrounded by a low-density shell, in which gravity is nearly balanced by turbulent motions ($\alpha\equiv K/|W|=0.4$). The clouds measure \( \num{3e3} \ \msun{} \) ($\num{3e4} \ \msun{}$) in mass and 4.6~pc (10~pc) in radius for the {\tt Ma7} ({\tt Ma15}) runs. The simulation box is 4 times larger than the diameter of the cloud to follow the expansion and dissolution of the cloud. We let the cloud relax for three free-fall times to allow the turbulence to develop before turning on full gravity and sink particle formation (\ie, star formation). Adaptive mesh refinement is applied to the whole domain to make sure at any time and any location the local Jeans length, $L_J = c_s \sqrt{\pi/(G \rho)}$, is resolved by at least 10 grid points. The maximum refinement level $l_{\rm max}$ is set to 14 in the whole domain. 

In the zoom-in simulations presented in this work, we rerun each GMC simulation starting right before the first sink particle (prestellar core) forms in the baseline run. \edit{We, therefore, define a ``core'' in our simulations as the progenitor of a sink particle or binary sink particles in the baseline simulation. In physical terms, ``cores'' are the largest isothermal spheres that form inside a GMC and are in hydrostatic equilibrium.} We define a ``zoom'' region, about 2~pc in size, at the location where the first core forms and set a higher refinement level of \( l_{\rm max} = 18 \) inside this region. 
\edit{We measure the core mass (Table~\ref{tab:1})  using two methods: (1) by applying a density threshold of $3000 \ \pcc{}$, and (2) by considering the mass of gas encompassed with a 0.1 pc radius. This approach is motivated by the fact that, during the isothermal phase, each core has a $\propto r^{-2}$ density profile that extends to a few $\times 10^5$ AU. Within this range, the mass enclosed is directly proportional to the radius enclosed.} To reach the best possible resolution with manageable computational power, we use a nested reﬁnement structure where $l_{\rm max}$ increases as it gets closer to the domain centre. The critical density for sink formation is $n_{\rm sink} = \num{3.6e9}~\pcc{}$ and $\num{7.7e8}~\pcc{}$ for the \texttt{Ma7} and \texttt{Ma15} clouds, respectively. 

\edit{In order to identify discs and study their properties in the zoom-in simulations, we work in a cylindrical frame of reference centred on the disc with $z-$axis parallel to the angular momentum of the overdense gas inside the core. Since discs are expected to be reasonably axis-symmetric, the density, velocity, pressure, and magnetic field are averaged on concentric rings and weighted by mass. 
We define a disc by employing two principal criteria: i) a density threshold and ii) rotational support. The density threshold is set at $10^8 \ {\rm cm^{-3}}$, a value limited by resolution but not too low to avoid large spiral arms. This threshold, roughly consistent with the literature \citep{Joos2012,Joos2013}, helps in distinguishing potential disc structures, which can manifest either as a flat, pancake-like shape, or a radiant shape with a thinner inner disc and thick outer part (as illustrated in Figure~\ref{fig:denprj}).  Alongside the density threshold, we assess rotational support through the Keplerianity parameter, $\beta_K \equiv v_{\phi} / v_{\rm kep}$. For a structure to be considered a disc, it has to have a disc-like geometry and be rotationally supported. The definition of ``rotational support'' is empirical. We impose the requirement $\beta_K \ge 0.5$, which is equivalent to a rotational-to-gravitational energy ratio $K_r/|W| = 0.5 \beta_K^2 \ge 0.125$. We further cut off the disc at a radius at which $\beta_K$ starts to drop to avoid including the puffy envelope. }

In order to measure the cloud magnetisation we use an alternative formulation of the dimensionless mass-to-flux $\mu$ given by Equation~(\ref{eq:mu}), which is more reliable for systems that depart from uniform density and physical symmetry. The following definition remains accurate for general inhomogeneous mass distribution and/or asymmetric geometry:
\begin{equation}\label{eq:eWeB}
 \mu^2 \equiv \frac{|{W}|}{{\cal B}} = \frac{18 \pi^2}{5} \frac{GM^2}{\Phi_B^2} = \frac{M^2}{M_{\Phi}^2}.
\end{equation}
Here, ${W}$ is the gravitational potential energy and ${\cal B}\equiv V B^2/(8\pi)$ is the magnetic energy.
The last equal sign holds for the uniform spherical geometry, in which case the $\mu$ defined in Equation~(\ref{eq:eWeB}) is equivalent to the mass-to-flux ratio definition. For a more centrally concentrated geometry, \eg{}, a non-singular isothermal sphere, the equivalent geometrical factor $c_{\Phi}$ is up to $70\%$ higher and $\mu$ is $40\%$ lower than the uniform density case (see Appendix~\ref{sec:mu}). 

\section{Results}
\label{sec:res}

\begin{table}
    \centering
    \begin{tabular}{lcccc}
         core name&   $R$ (AU)& $H$ (AU)&$t$ (kyr)&$N_{\rm sink}$\\
         \hline
         $\mu$3Ma15& 
     1000&  400&230&6\\
 $\mu$3Ma15-large& 5000&  3000&>500&9\\
 $\mu$1Ma15& 300&  100&>60&2\\
 $\mu$0.6Ma15& None&  None&None&32\\
 $\mu$3Ma7& 600&  200&>100&3\\
 $\mu$3Ma7-hires& 600&  200&>100&12\\
 $\mu$1Ma7& 1000&  200&>10&21\\
 $\mu$0.6Ma7& None&  None&None&112\\\end{tabular}
    \caption{\edit{disc properties from simulations listed in Table~\ref{tab:1}. The columns are core names, disc radius (in AU), disc thickness (in AU), disc lifetime (in kyr), and number of sink particles formed in the vicinity of the disc that are either orbiting, ejected, or at the centre of the system.} }
    \label{tab:2}
\end{table}

\edit{The key result from our suite of simulations is that large, Keplerian circumstellar discs form in magnetically (near-)critical cores. We present the disc properties and sink particle statistics within the discs in Table~\ref{tab:2}. We refer to our companion paper of this series (Paper~I) for more details on the sink mass functions and a detailed study on disc properties. } In this paper we focus on the role of the large scale magnetic field in the GMC in determining the discs formation and properties. Below we present results, for a set simulations with different $\mu$ and turbulent Mach number, focusing on the cores morphologies and evolution (\S~\ref{sec:mor}), the magnetic field-gas density relationship (\S~\ref{sec:b-rho}), the vertical support of the discs (\S~\ref{sec:b-rho}), and a critical analysis of the magnetic breaking problem (\S~\ref{sec:braking}). 

\subsection{Morphology of Gas Density and Magnetic Field in Cores}
\label{sec:mor}
\begin{figure*}
  \newcommand\gap{\hspace*{-0.9cm}}
  \newcommand\ww{1.9in}
  \newcommand{\textbox}[1]{\framebox[1.6in]{#1}}
  \setlength{\fboxrule}{0pt}
  \centering
  \textbox{$\mu$3Ma15}\textbox{$\mu$1Ma15}\textbox{$\mu$1Ma7}\textbox{$\mu$0.6Ma15}\\
  \includegraphics[width=\ww]{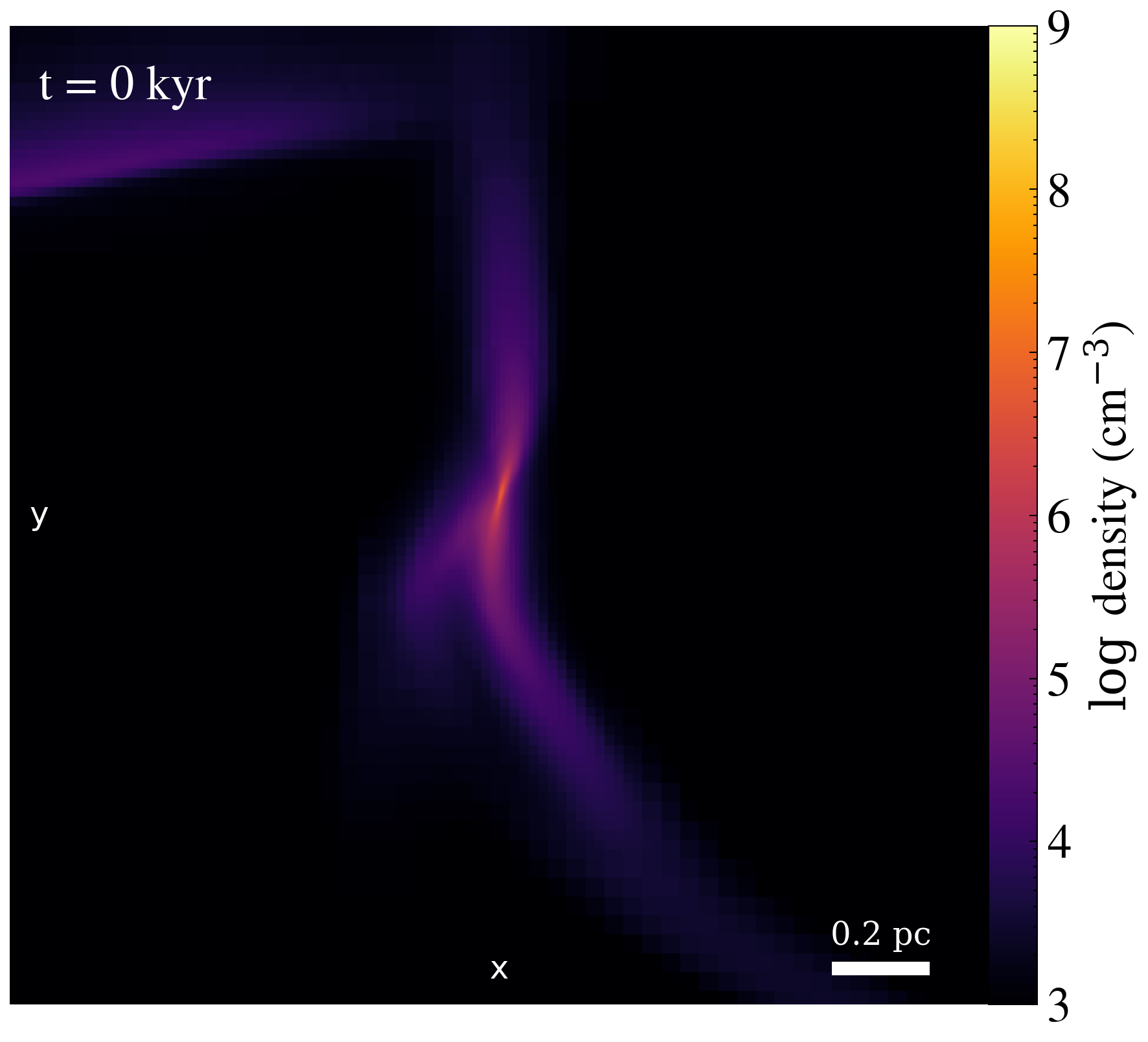}
  \gap{} \includegraphics[width=\ww]{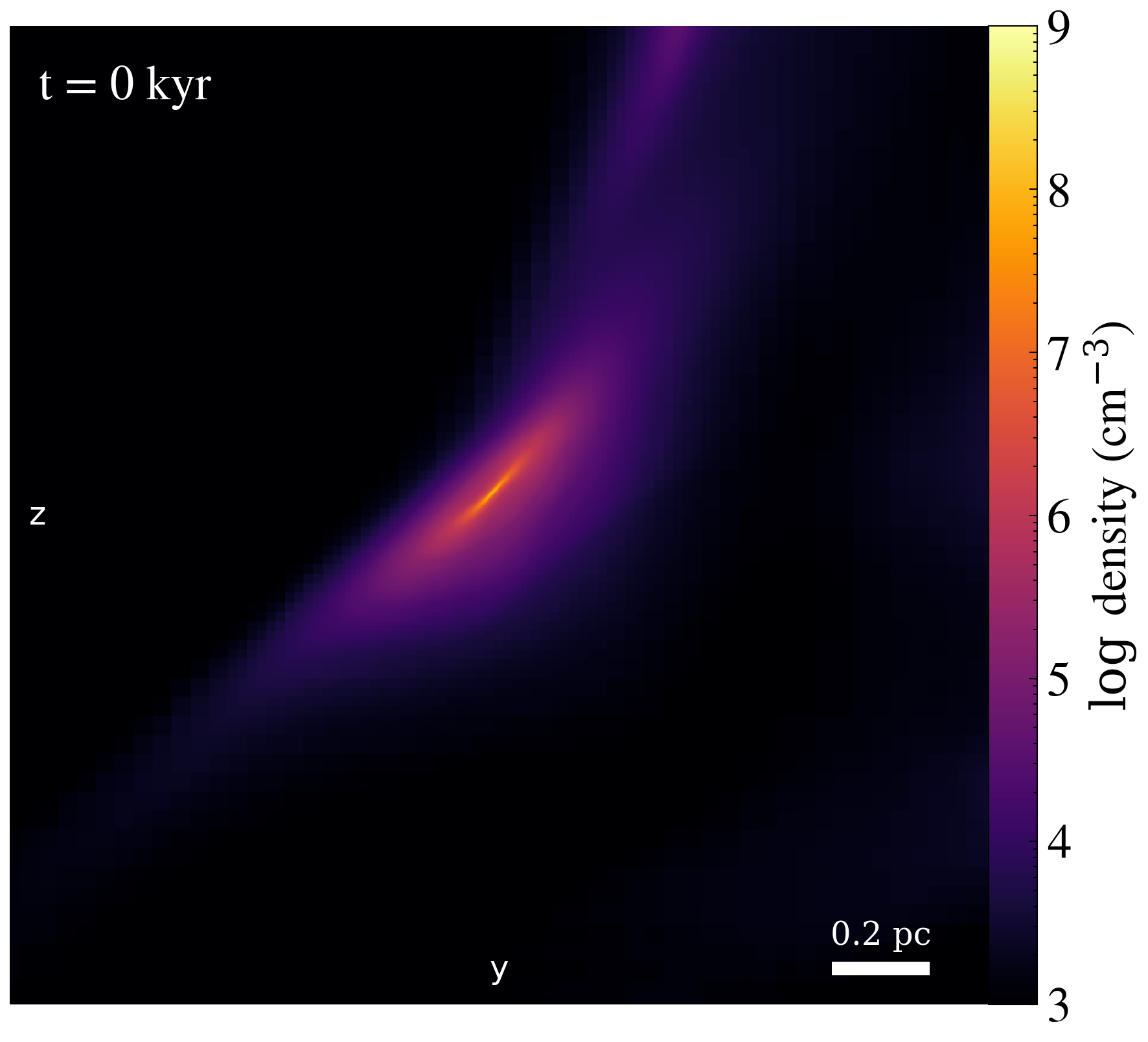}
  \gap{} \includegraphics[width=\ww]{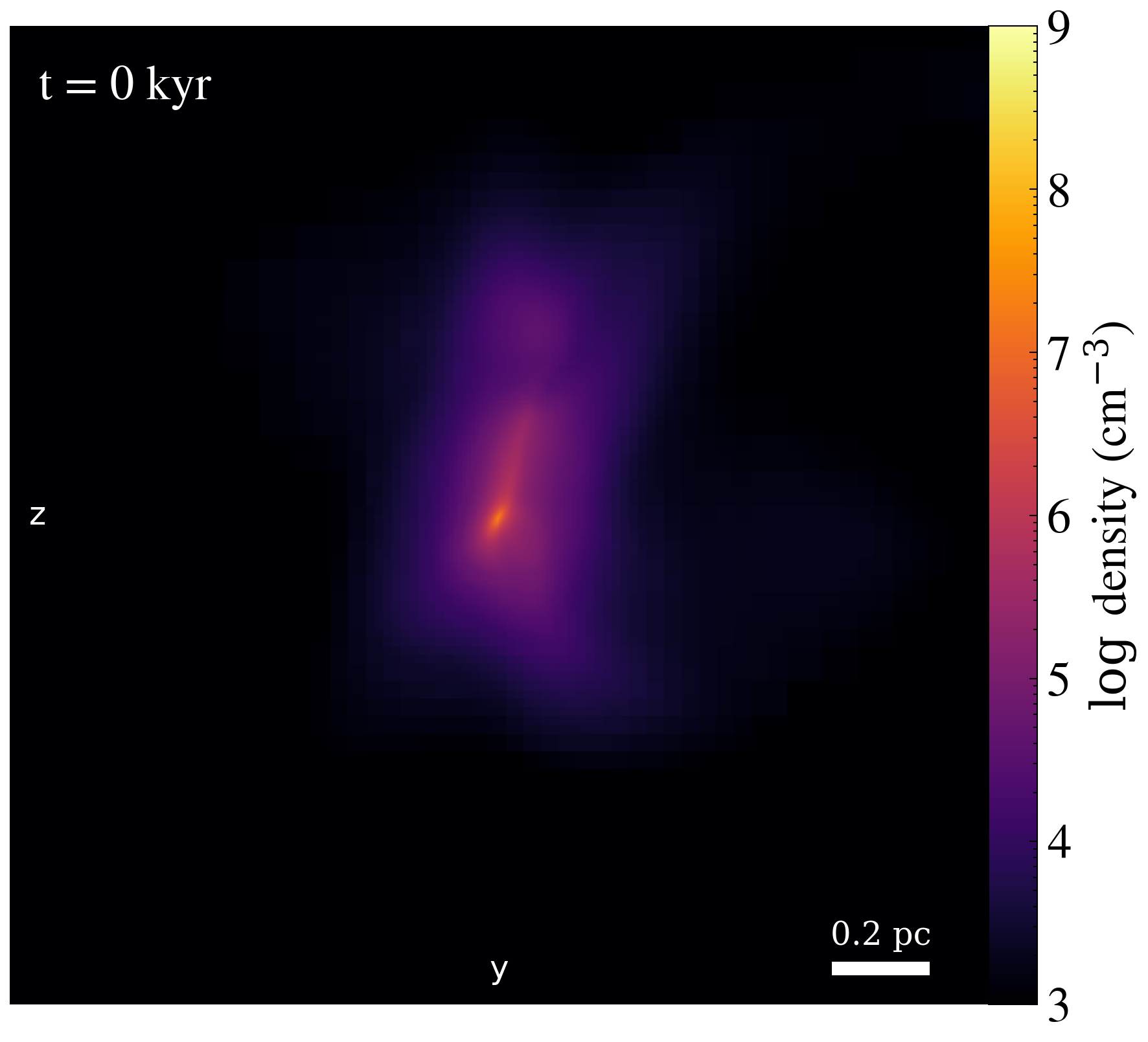}
  \gap{} \includegraphics[width=\ww]{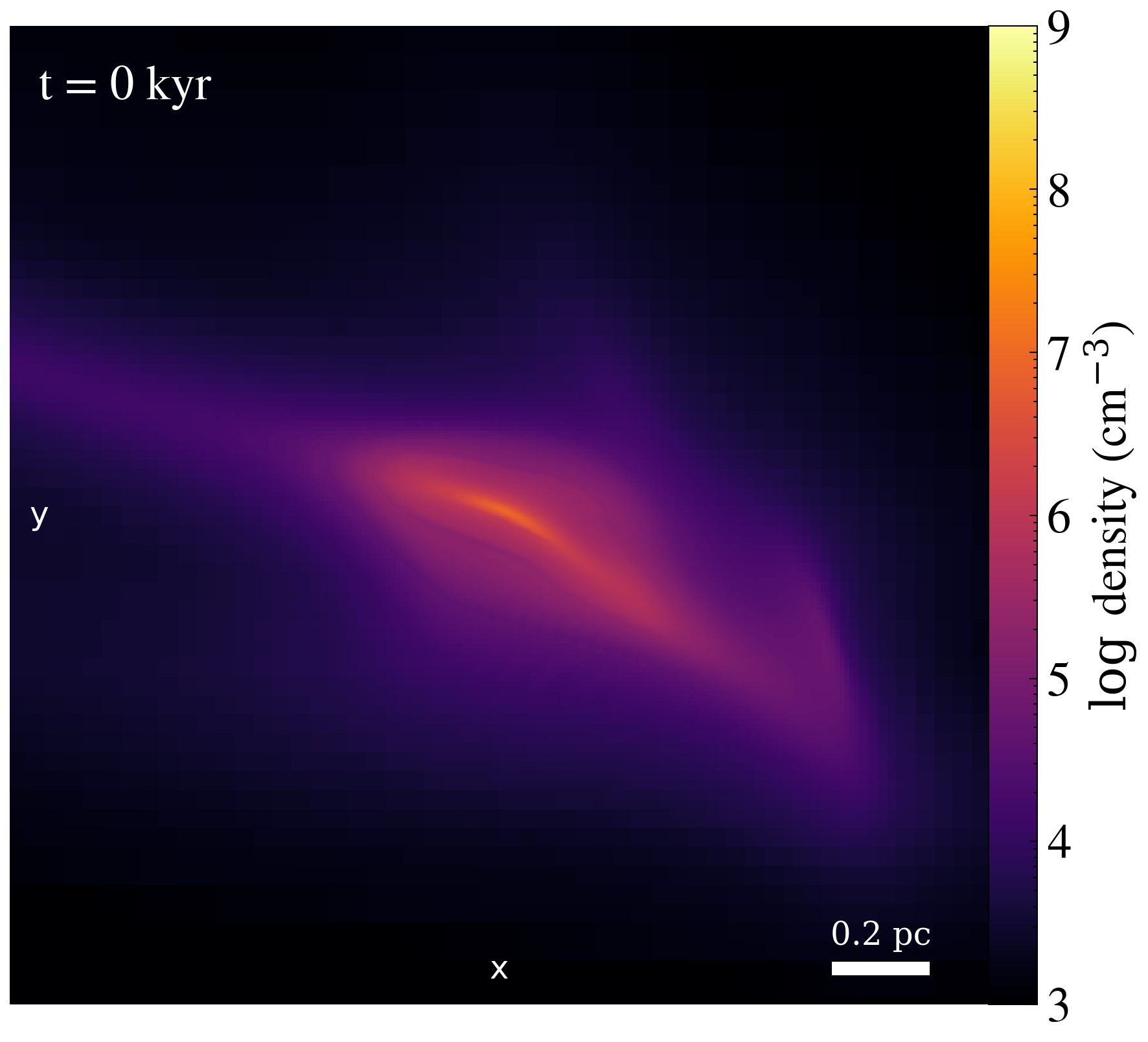}
  \\ \vspace{-0.2cm} 
  \includegraphics[width=\ww]{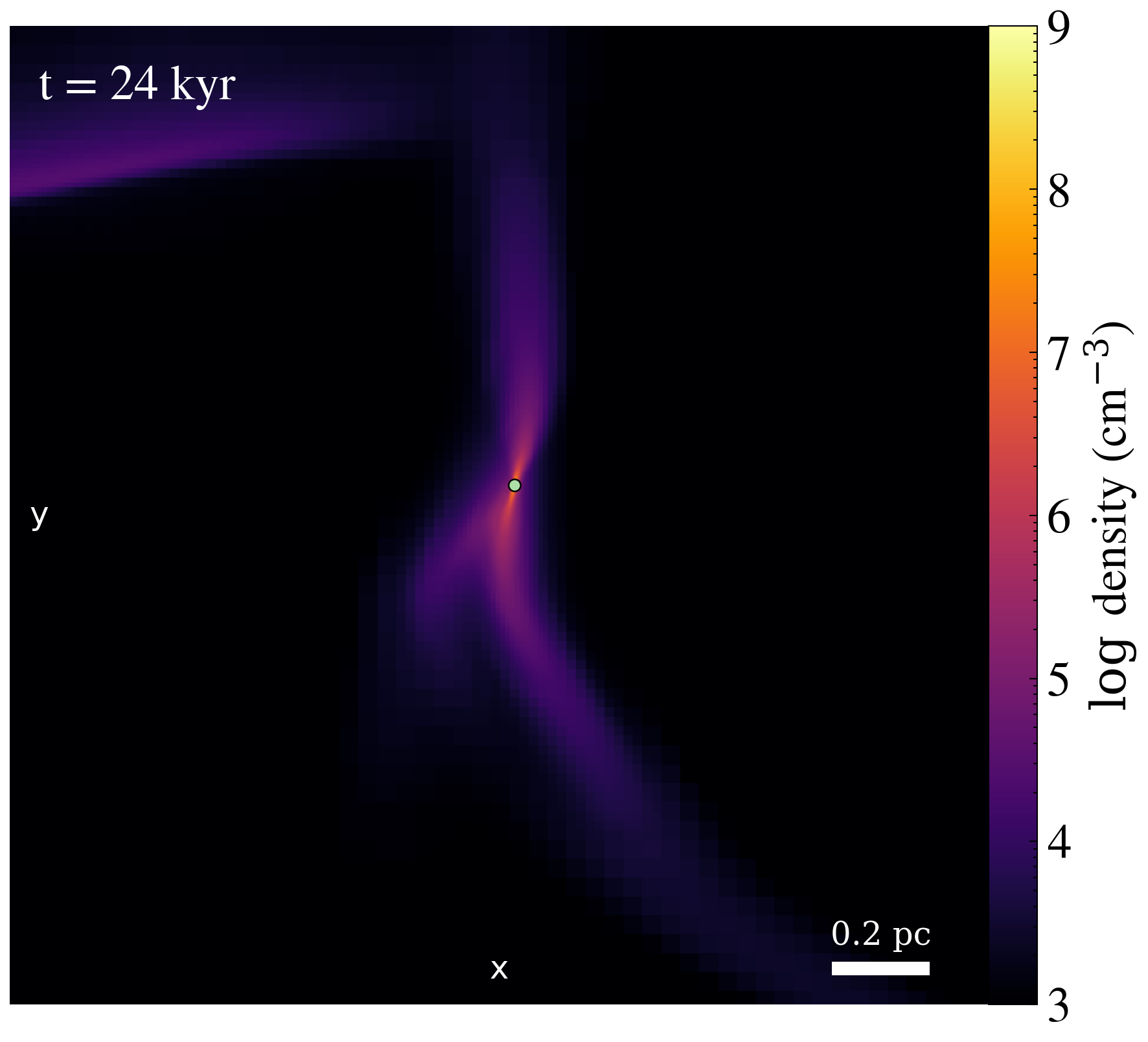}
  \gap{} \includegraphics[width=\ww]{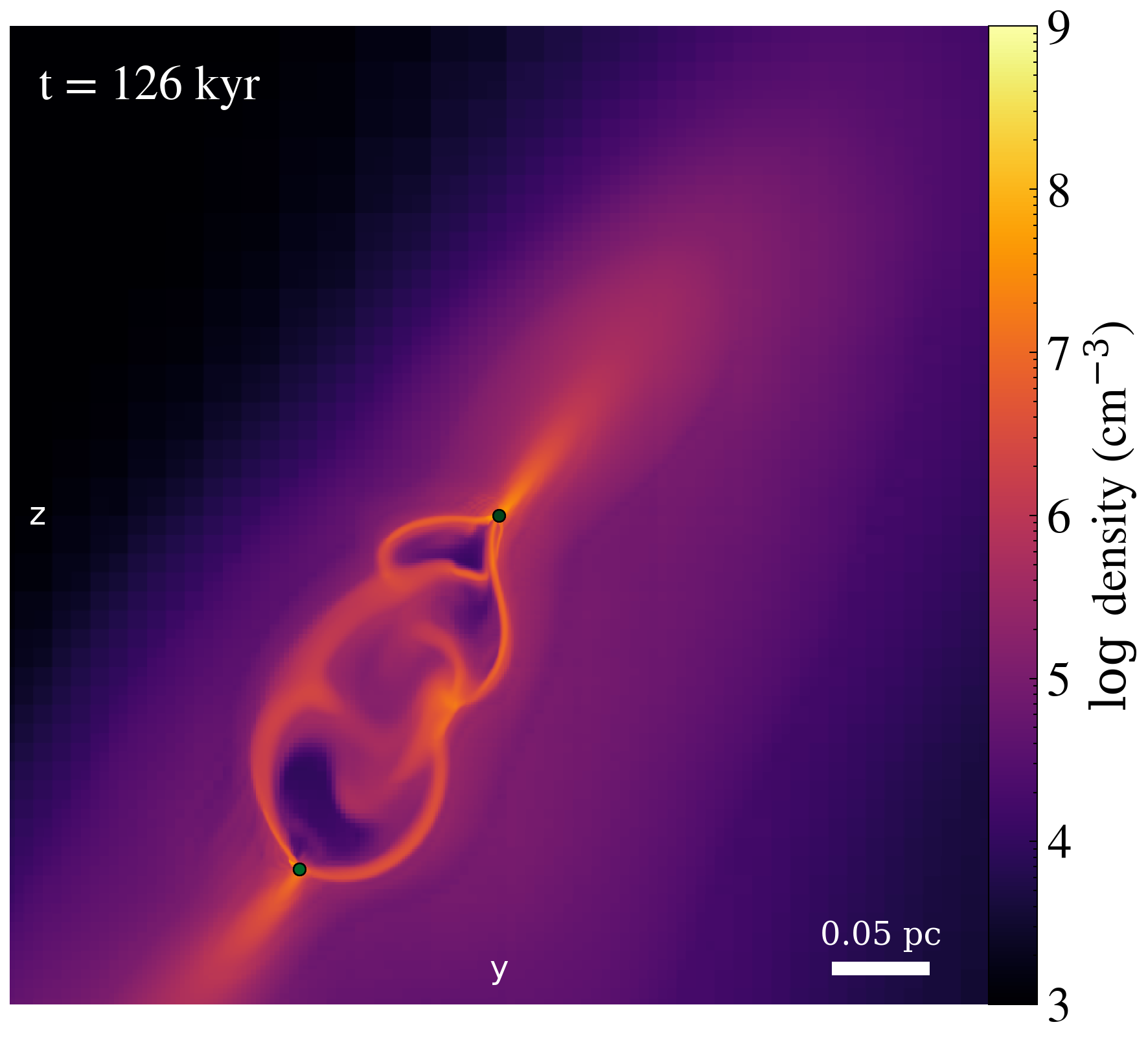}
  \gap{} \includegraphics[width=\ww]{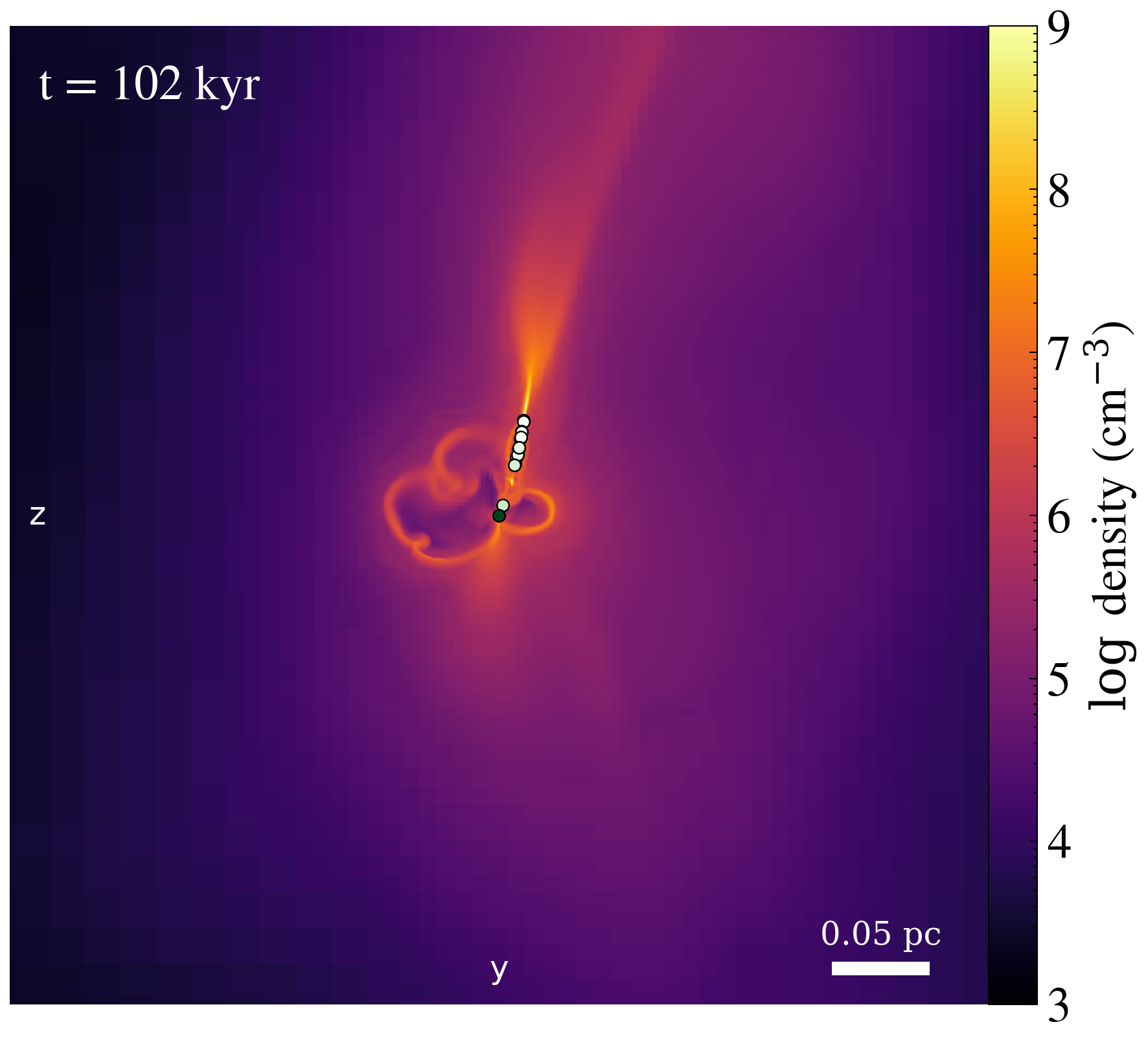}
  \gap{} \includegraphics[width=\ww]{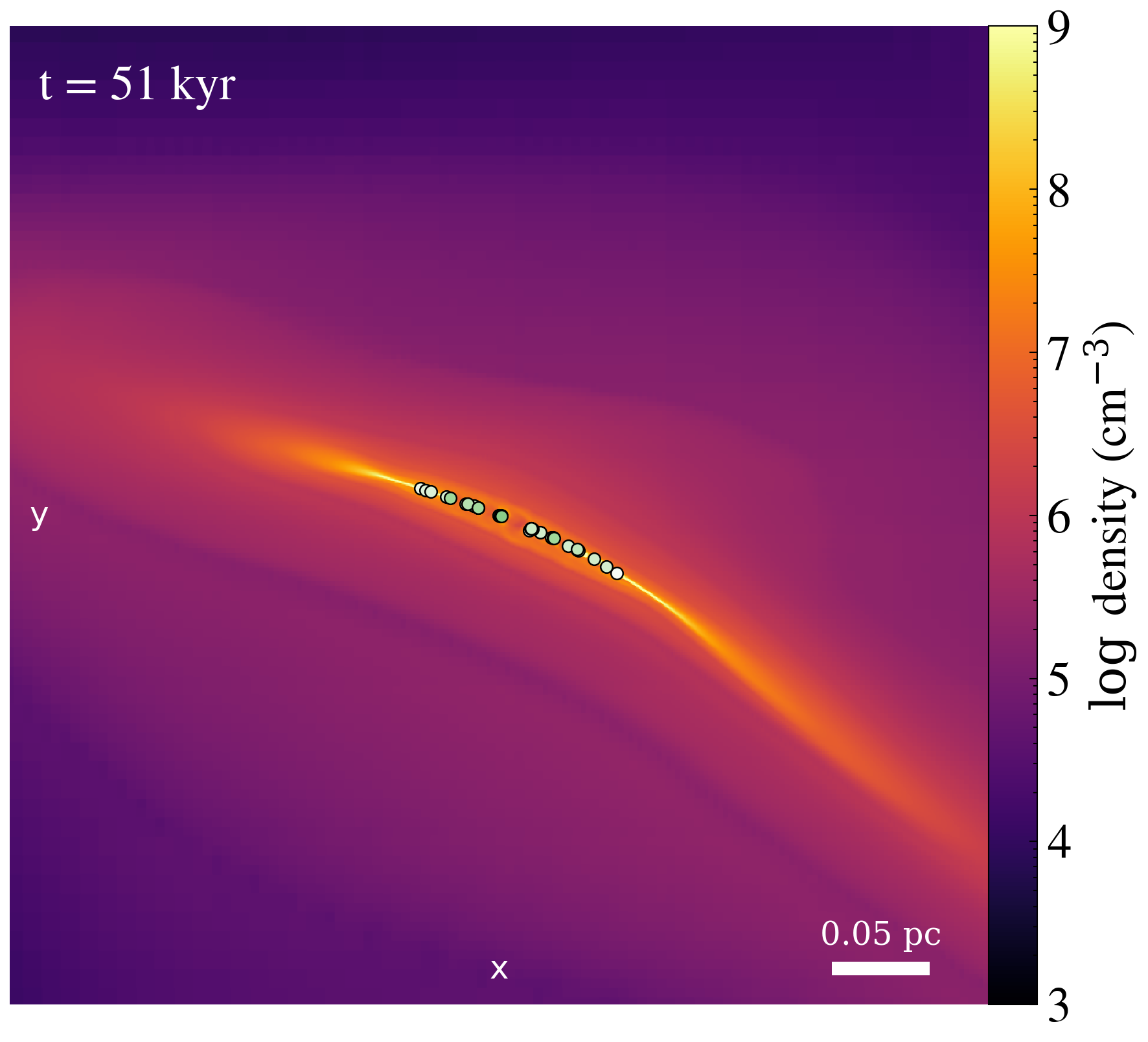}
  \\ \vspace{-0.2cm}
  \includegraphics[width=\ww]{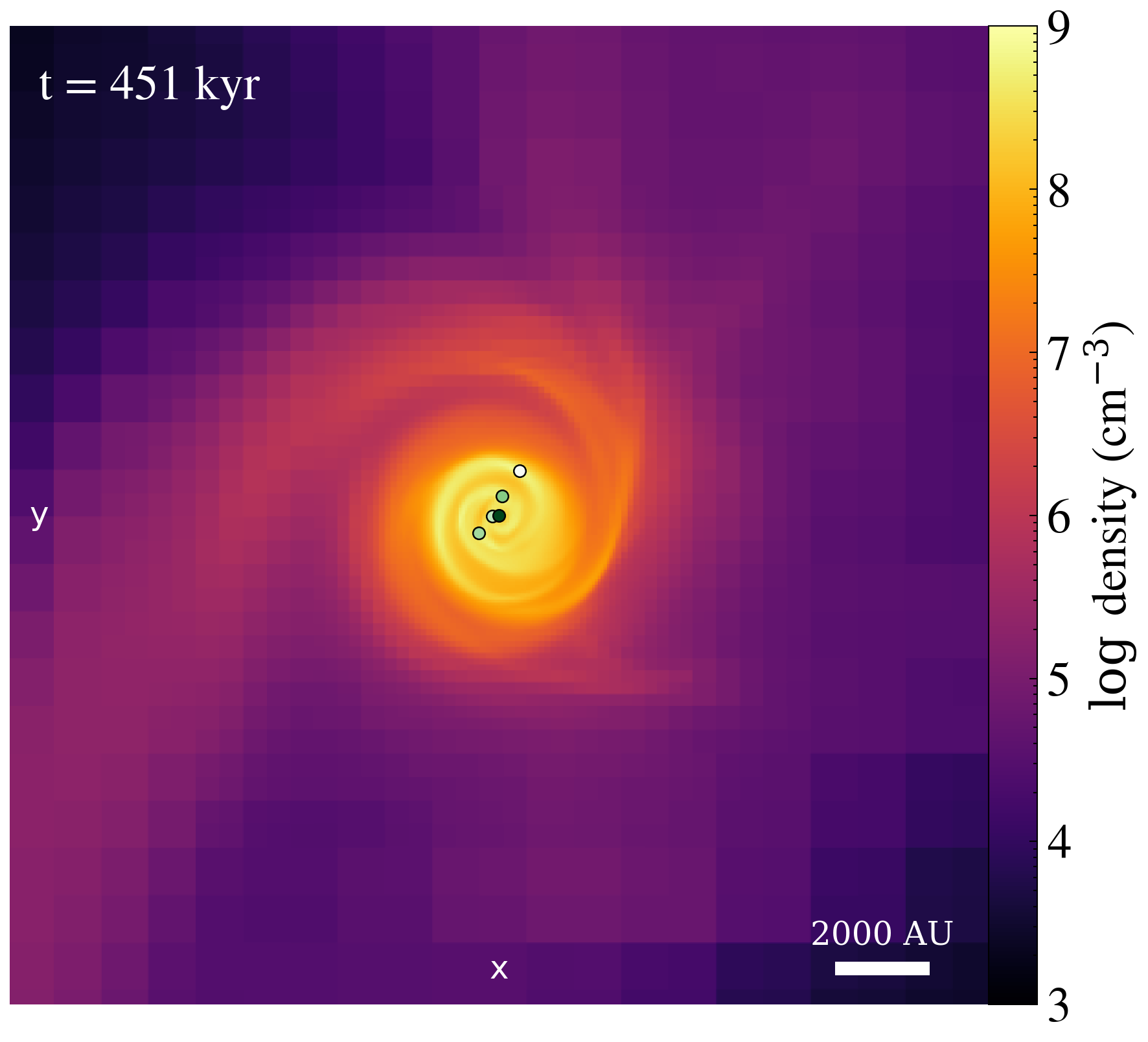}
  \gap{} \includegraphics[width=\ww]{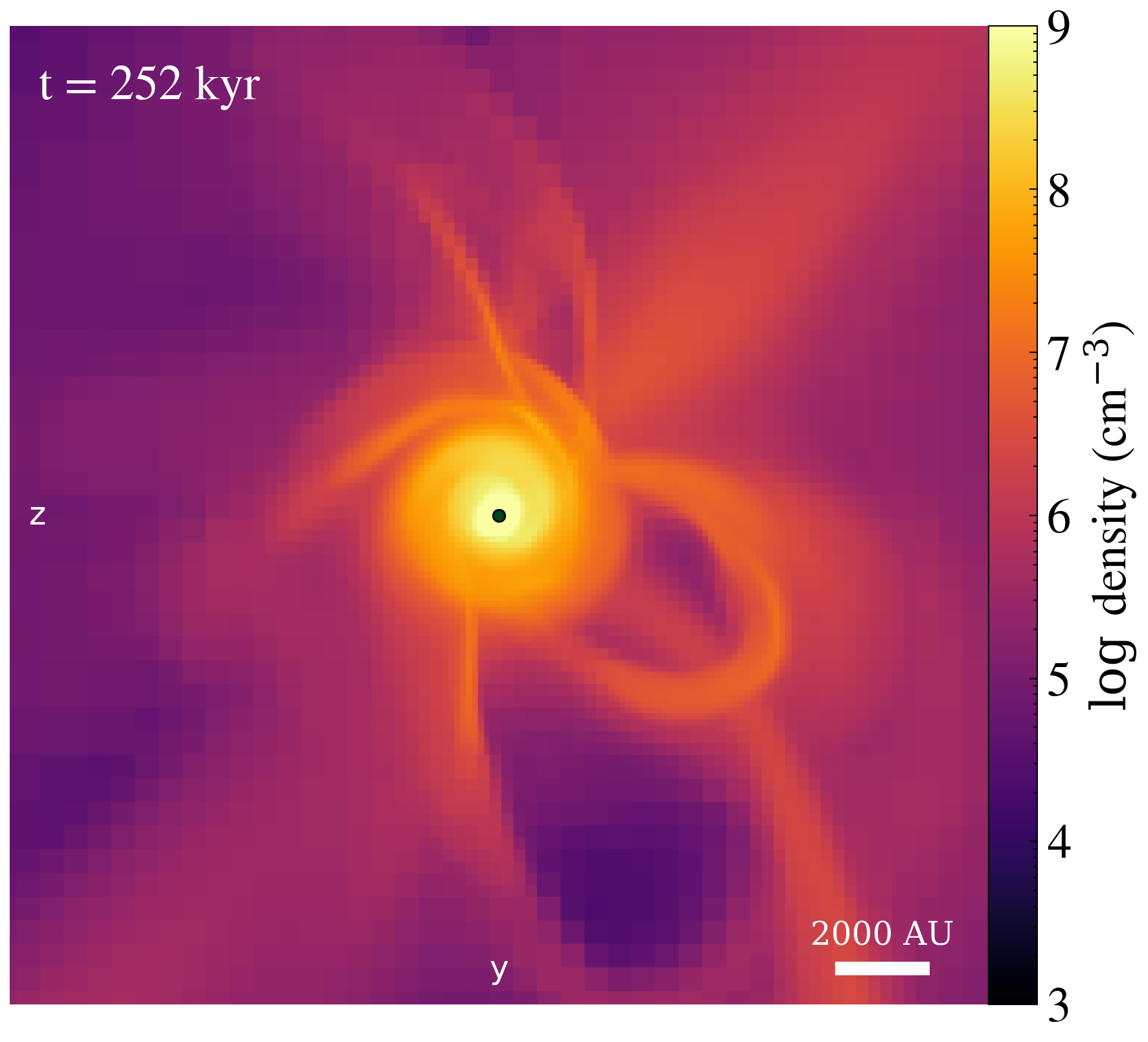}
  \gap{} \includegraphics[width=\ww]{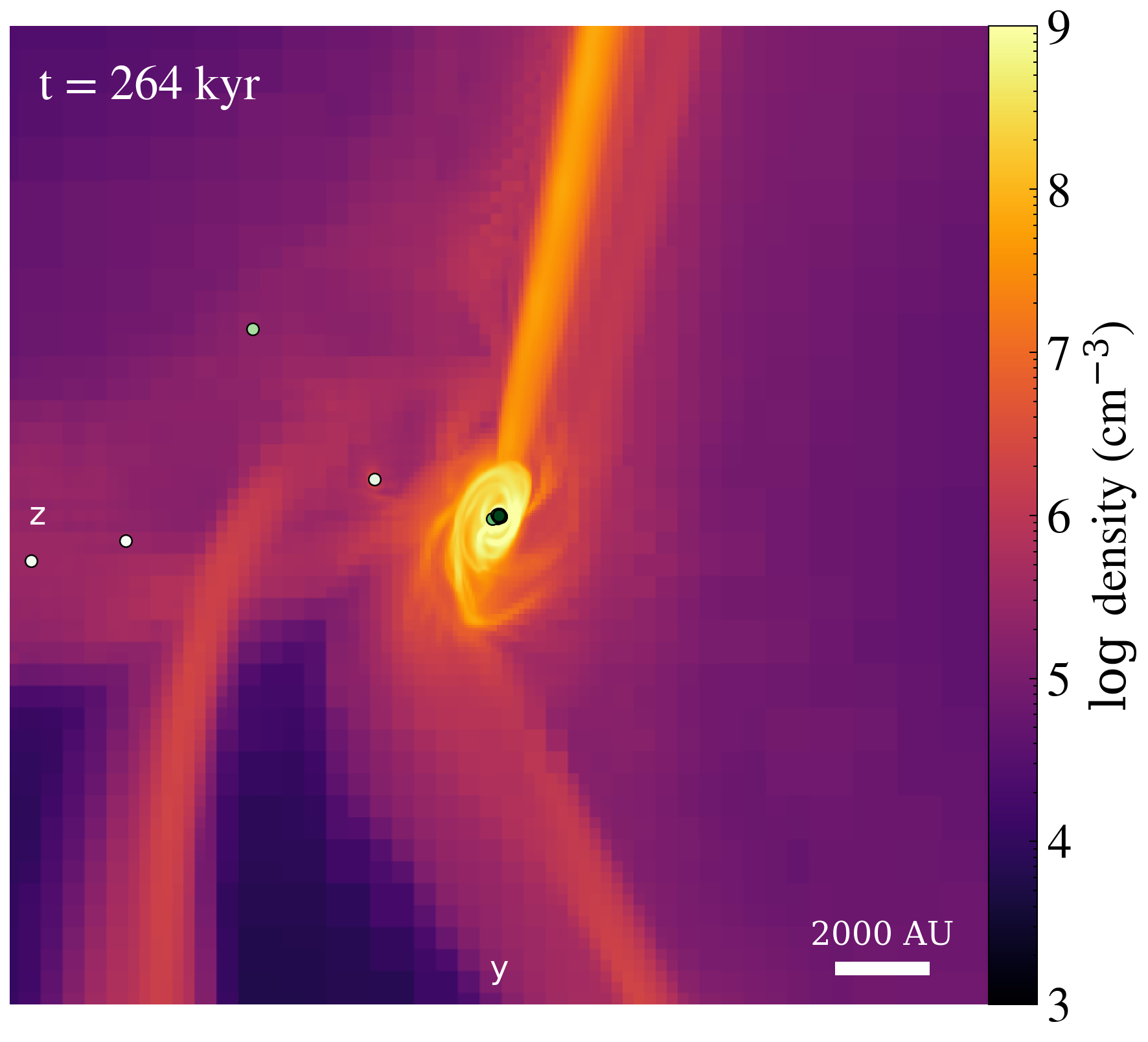}
  \gap{} \includegraphics[width=\ww]{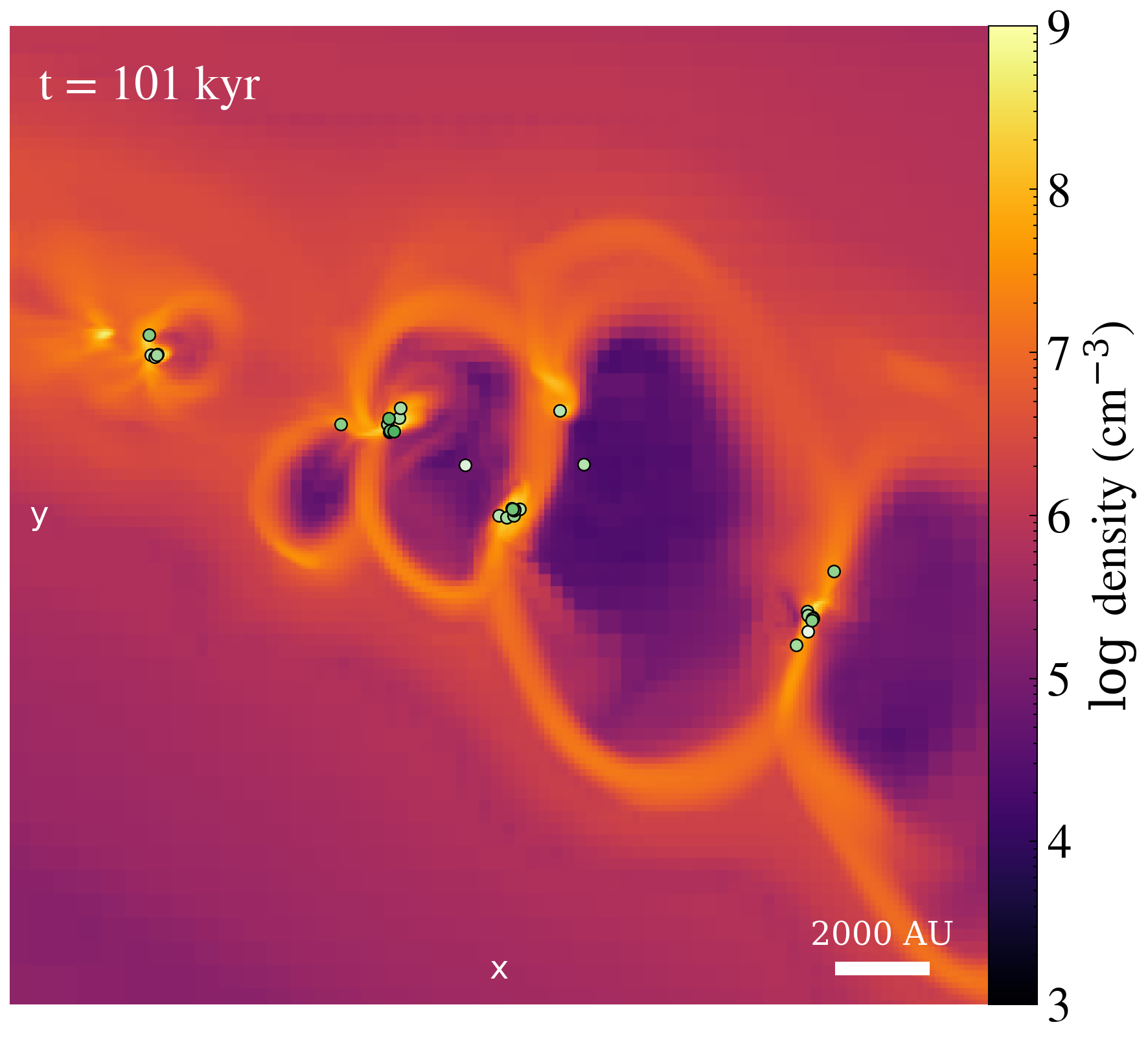}
  \caption{\label{fig:denprj} Density projections of the four cores (from left to right) inspected in Figure~\ref{fig:1}, showing their time evolution (from top to bottom) and the formation of stars. The circles mark the positions of the sink particles which represent individual stars. The colours of the circles, from white to dark green, correspond to their masses from 0.1 \msun{} to 10 \msun{} in log scale. 
  A large Keplerian disk forms in all cores in GMCs with $\mu_0 \gtrsim 1$. In the case where $\mu_0 < 1$ (\cLH{}), the core collapses into a sheet-like shape with negligible angular momentum with respect to the centre of mass, preventing disc formation.
  }
\end{figure*}  

For the same GMC simulation and sink particle we zoom into, the strength of the background magnetic field has an effect on the early phase of collapse and the initial morphologies of the collapsing cores. The top row of Figure~\ref{fig:1} shows the surface density of the gas for a representative set of 4 simulations while to bottom row shows the density in a slice through the same filament and the magnetic field lines. In the weaker magnetic field cases (\cHH{}, \cMH{}, and \cML{}), the core collapses starting from a gas filament oriented perpendicular to the magnetic field lines. This nearly one-dimensional structure is denser near its centre of mass and the density decreases in the outer parts of the filament. Both the projection plots and the slice plots show similar morphology and orientation of the filament geometry, which is also confirmed by an inspection of the 3D rendering of the gas density. In the stronger magnetic field case (\cLH{}), the gas also collapses along the field lines but into a two-dimensional sheet-like structure. The gas surface density of the sheet is also not uniform and the peak density defines a one-dimensional curve, or filament, that is not necessarily oriented perpendicular to the B-field lines. This can be observed in the projection plot for run \cLH{} in Figure~\ref{fig:1}, showing a filament structure apparently oriented along the B-field lines. However, the slice plot indeed shows the magnetic field lines threading through the sheet are perpendicular to its surface.
To summarize, since the surface density is the quantity more readily observable, in the weak B-filed case the surface density maps reflect the actual filamentary shape of the gas leading to core formation. In the strongly magnetised cloud, however, the surface density map may lead observers to mistakenly think that the structure is a filamentary or disc-like structure rather than a two-dimensional sheet structure. 

\begin{figure*}
  \centering
  \includegraphics[width=0.95\textwidth]{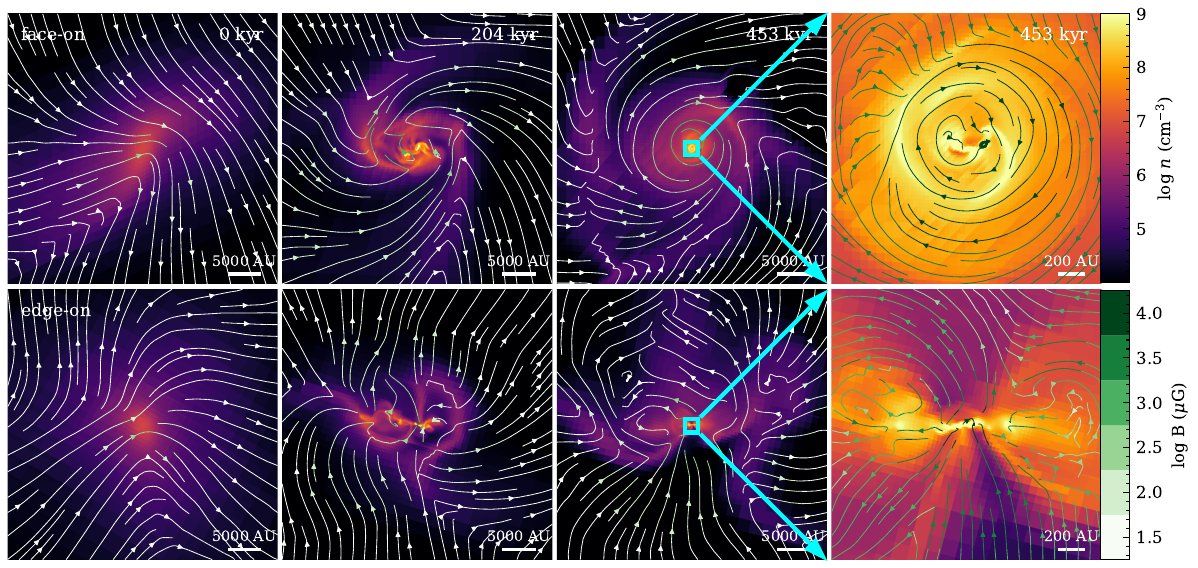}
  \caption{Temporal evolution (from left to right) of the magnetic field morphology of core {\tt $\mu$3Ma15-hires} in a face-on view (top row) and edge-on view (bottom row). Each panel shows a slice of the gas density centred at the peak density overplotted with magnetic field streamlines. The colour of the streamlines shows the magnetic strength as indicated by the colorbar at the bottom right. 
  Note that the last column shows a zoomed view of the centre of the disc for the same time snapshot shown in the third column panels. These figures show how the magnetic field lines are wound up as the disc forms and that the magnetic field is extremely turbulent, especially near the disc centre.}
  \label{fig:stream}
\end{figure*}

\begin{figure}
  \centering
  \includegraphics[width=3.2in]{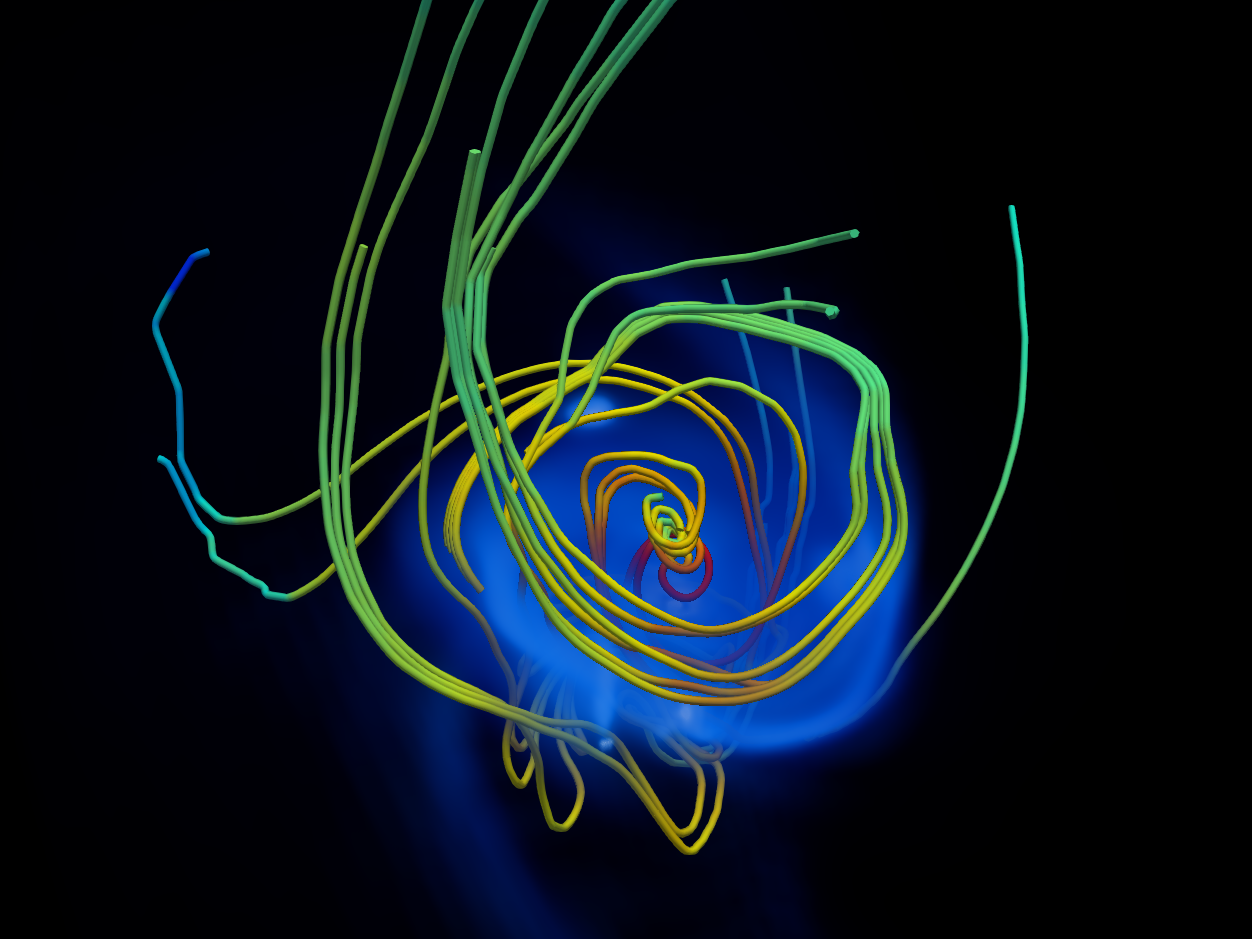}
  \caption{3D view of the magnetic field lines on top of the volume rendering of the disc {\tt $\mu$3Ma15-hires}. The rendering in blue shows the \edit{disc envelope with} densities above $10^6~\pcc{}$. \edit{Note that the disc itself is at the centre and is much smaller.} The tubes indicate the direction of the field lines and their colours indicate the strength of the magnetic field with red to blue meaning strong to weak. \edit{The field of view measures 10,000 AU. }}
\label{fig:mag3d}
\end{figure}

During the next stage of collapse, the gas fragments along the filament or along the central high-density line of the sheet (Figure~\ref{fig:denprj}). In the supercritical/critical ($\mu_0 \ge 1$) cases, the core fragments into multiple stars along the filament. Due to the conservation of angular momentum, the filament spirals inward and forms a Keplerian disc. The fragments are ultimately embedded in the Keplerian disc, as already shown in our previous work (Paper~I). In the cases where $\mu_0 < 1$, the sheet-like structure collapses toward the central high-density line. When the central density reaches the local Jeans density, the gas collapses to form numerous stars that are clustered into clumps. Due to magnetic braking, as will be discussed in \S{}~\ref{sec:braking}, the angular momentum of the gas is transferred outward and the gas flows directly into the centre of the local clump without forming a disc. 

To take a peek at how the magnetic field morphology evolves over time, we plot the magnetic field lines of core \texttt{$\mu$3Ma7-hires} at four snapshots in Figure~\ref{fig:stream}. As the cores collapse, the magnetic field is bent by the rotation of the gas through the dynamo effect. The field lines are twisted and the magnetic strength is enhanced at the disc centre, which is also demonstrated by the 3D volume rendering of the magnetic field lines in Figure~\ref{fig:mag3d}. The field is dominated by the toroidal component in the inner region and by the poloidal component in the outer region, which has implications on the strength of magnetic braking as will be discussed in \S{}~\ref{sec:braking}.

\begin{figure*}
  \centering
  \includegraphics[scale=0.72]{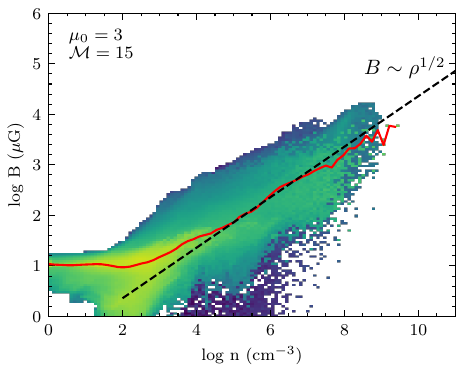}
  \includegraphics[scale=0.72]{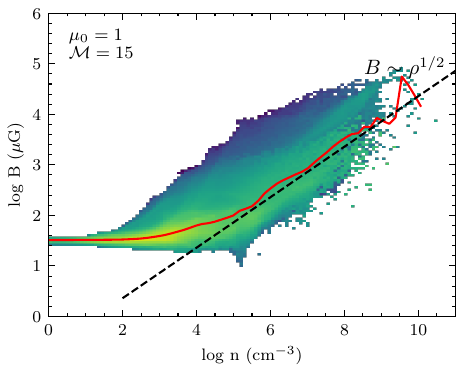}
  \includegraphics[scale=0.72]{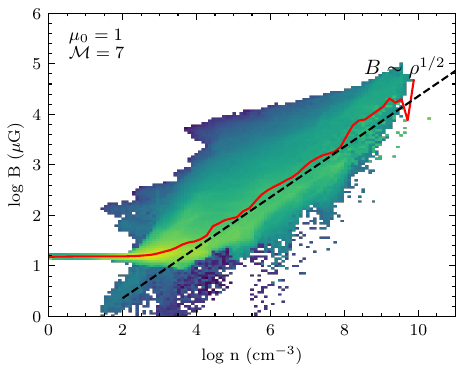}\\
  \includegraphics[scale=0.75]{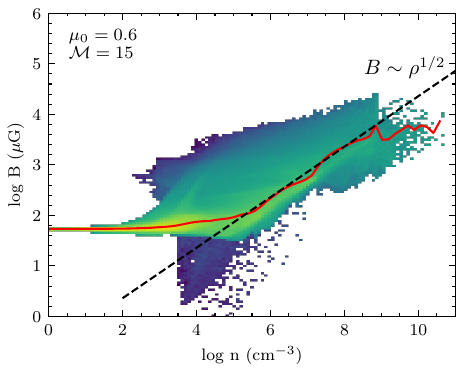}
  \includegraphics[scale=0.75]{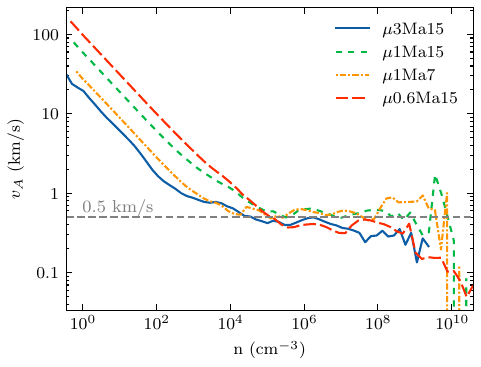}
  \caption{The global $B-\rho$ relationship at GMC scales for the \cHH{}, \cMH{}, \cML{}, and \cLH{} runs. At low density ($\lesssim 10^3 \ \pcc$), the magnetic field intensity is independent of the density and the global value is determined by the initial magnetization. At high density up to $\sim 10^9 \ \pcc$, we find a universal $B-\rho$ relation: $B \approx 86 \ \mu\textrm{B} \ (n/10^5 \ \pcc{})^{1/2}$, corresponding to a constant Alfven velocity of $\sim$ 0.5 km/s.
    The first four panels show the 2-D phase diagram of magnetic intensity $B$ vs gas number density $n$ of the four clouds, respectively, and the colours represent the log of the gas mass with increasing mass from dark to bright. The red curves are the mass-weighted 1-D relationship. The last panel plots the Alfven velocity $v_A$ as a function of $n$, converted from the 1-D $B-n$ relationship for all four clouds. 
}
  \label{fig:brho1}
\end{figure*}

\begin{figure*}
  \centering
  \def\thew{0.32}
  \includegraphics[width=\thew\textwidth]{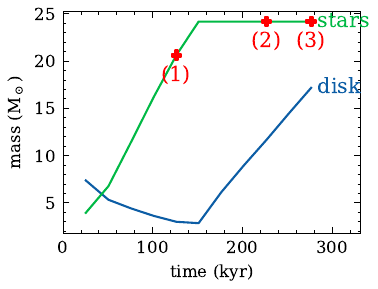}
  \includegraphics[width=\thew\textwidth]{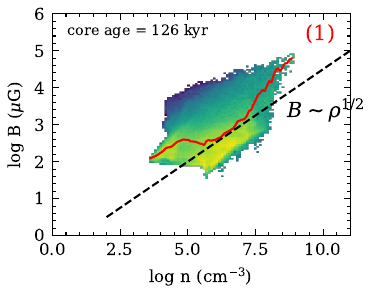}\\
  \includegraphics[width=\thew\textwidth]{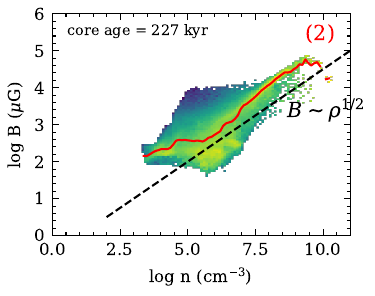}
  \includegraphics[width=\thew\textwidth]{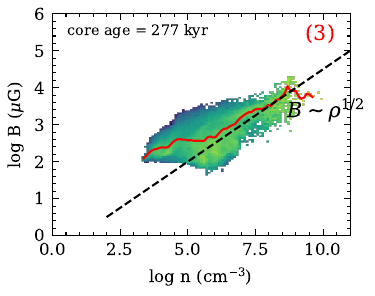}
  \caption{Enhancement of the magnetic intensity at high density inside the core \cMH{}. The first panel shows the temporal evolution of the mass in stars and in discs. The rest of the panels show a 2-D phase diagram of $B-n$, similar to Figure~\ref{fig:brho1}, inside the core at various times as marked in the first panel. We show that as the gas is accreted by sink particles while the magnetic field is retained outside, the magnetic intensity is boosted by nearly an order of magnitude at a density above $10^7 \ \pcc{}$. As the accretion stops, the strong magnetic field disperses itself due to magnetic pressure/tension.}
  \label{fig:brho2}  
\end{figure*}

\subsection{\texorpdfstring{$B-\rho$}{B-rho} Relationship}
\label{sec:b-rho}

Understanding the relationship between density and magnetic field, as well as their physical origin, is crucial for both observations and theoretical models. According to simple models \citep[\eg,][]{Crutcher1999}, the magnetic field and gas density of a collapsing molecular cloud follow a power-law scaling relationship, $B \sim \rho^{\kappa}$. Assuming flux-freezing in ideal MHD, in the scenario of the isotropic collapse of a spherical cloud threaded by uniform parallel magnetic field lines, the relationships $B \propto R^{-2}$ and $\rho \propto R^{-3}$, imply $B \propto \rho^{2/3}$.
In the scenario of the anisotropic collapse of a flattened structure or disc, the evolution of the gas collapse consists of two stages: during the first stage the cloud collapses along the field lines to form a disc. The gravitational acceleration at the disc surface, according to Gauss's law, is approximately $g \approx -2 \pi G \Sigma$, where $\Sigma=\rho H=M/(\pi R^2)$ is the gas surface density, and $H$, $R$ and $M$ are the disc thickness, radius and mass, respectively. In the second stage, assuming flux-freezing, the mass-to-flux ratio $M / \Phi_B = \Sigma/B$ is conserved, therefore $B \approx \Sigma \Phi_B / M$. Assuming $g \approx \sigma^2 / H$, \ie, the disc is supported by turbulent pressure in the vertical direction, we find $\sigma^2 \approx 2 \pi G \rho H^2 \sim 2\pi G \Sigma^2/\rho$. Therefore,
\begin{equation}\label{eq:1}
    B \approx \frac{\sigma}{\sqrt{2 \pi} \ c_\Phi \mu} \rho^{1/2},
\end{equation}
or expressed in terms of the Alfven velocity
\begin{equation}\label{eq:va1}
  v_A = \frac{B}{\sqrt{4 \pi \rho}} \approx \frac{\sigma}{2 \sqrt{2} \pi c_\Phi \mu}.
\end{equation}
An even simpler interpretation of this relation is the equipartition between magnetic and kinetic energy, $B^2/(4\pi) \propto \rho \sigma^2$, noting that
\begin{equation}
  \frac{v_A^2}{\sigma^2} = \frac{1}{8 \pi^2 c_\Phi^2 \mu^2} = \frac{0.45}{\mu^2}. 
  \label{eq:va}
\end{equation}
In flattened cores where $\mu$ is marginally greater than 1, this relationship predicts that magnetic pressure is slightly weaker than turbulent pressure.

\edit{Our discussion of the $B-\rho$ relationship is organised into two regimes: the GMC scales and core scales. The former is the larger scale, encompassing all gas inside the simulation box that houses the GMC. The latter focuses on the more localized region of prestellar cores, specifically targeting the gas contained within an individual core.}

\subsubsection{$B-\rho$ Relationship in GMCs}

It is well established \citep{Troland1986,Crutcher2010} that at low densities ($\lesssim 10^3 \ \pcc{}$), the intensity of the magnetic field does not depend on gas density. This phenomenon is clearly seen in our simulations (Figure~\ref{fig:brho1}), and the intensity is determined by the strength in the initial condition. At higher densities, up to $\sim 10^9 \ \pcc$, the magnetic field scales with density as $B \propto \rho^{1/2}$, resulting in a universal value of the Alfven velocity of $v_A \sim 0.5\pm 0.1$~km/s. 
In Figure~\ref{fig:brho1} we show phase plots of $B$ vs gas number density $n$ for GMCs with a range of $\mu_0$ from $3$ to $0.6$ and for turbulent Mach numbers ${\cal M}=7$ and $15$. The red line, showing the \edit{mass-weighted average} of $B$ at a given $n$, suggests that the Alfven velocity derived from this median relationship is nearly constant independently of $\mu_0$ and ${\cal M}$.
It is important to note, however, that the Alfven velocity is only approximately constant above the critical density, and the exact value may also depend on other properties of the cloud. In Section~\ref{sec:intBrho}, we will explore the physical mechanisms that govern the scaling of the magnetic field with density in more detail.
The critical density at which there is a transition from $B=B_0={\rm const.}$ to $v_A={\rm const.}$ is:
\begin{equation}
n_{\rm cr} \equiv \frac{1}{4\pi \mu_{H}}\left(\frac{B_0}{v_A}\right)^2 \approx \num{2.2e4}~{\rm cm}^{-3}\left(\frac{B_0}{30~\mu G}\right)^2.
\end{equation}
This critical density is roughly consistent with the observed density of cores in GMCs, suggesting that cores form when the magnetic field is no longer strong enough to support the cloud against gravitational collapse.

\subsubsection{$B-\rho$ Relationship in Cores}

In Figure~\ref{fig:brho2}, we show the $B-\rho$ relationship for gas in the cores of the zoom-in simulations. \edit{In the cores, the gas with density lower than $10^7 \pcc{}$ approximately follows the same universal relationship $B \propto \rho^{1/2}$ with the same normalisation as at GMC scales (Figure~\ref{fig:brho1}).} 
However, at densities $>10^{7}$~\pcc we notice that the magnetic field intensity at a given density is enhanced with respect to the universal value when the sink particle accretes gas. This phenomenon is likely a result of the release of the magnetic field during sink accretion, as the sink particles accrete gas but not magnetic field, hence the conservation of mass-to-flux ratio, valid in ideal MHD, is broken within the sink particles. This recipe for sink accretion is particularly motivated by the magnetic flux problem in star formation: the mass-to-flux ratio in a star is $10^{5-8}$ times higher than that at the cores' scale. \edit{The slope of the $B-\rho$ relation at densities $>10^{7}$~\pcc is close to $2/3$, suggesting that the collapse at disc centre is nearly isotropic}. The boost in the $B$ field at a given $\rho$ persists as long as the sink particle is accreting gas. Shortly after the accretion stops, the magnetic field strength reduces back to the average value following the global $B-\rho$ relationship. Evidently, the accumulation of B-field lines in the sink produces a temporarily stronger magnetic pressure diffusing the magnetic field lines outward \edit{ \citep{Zhao2011}.}

\subsubsection{Interpretation of the Universal B-$\rho$ Relationship}
\label{sec:intBrho}

In all the phase plots we observe two regimes: (A) a low-density regime where the mean of the magnetic field is constant as a function of $\rho$, even though the spread around the mean can be large, especially for weaker values of the initial B-field (large $\mu$ cases); (B) a high-density regime, where $B\propto \rho^{1/2}$, or $v_A={\rm const}(\rho)$.
These two regimes are observed for the GMC as a whole (in this case the constant $B$ value is the one set in initial conditions), and for individual cores: in this second case the constant $B$ value regime is the one at the boundary of the core where the density is lowest.

The regime (A) is the case when the density of the gas can increase or decrease while leaving $B$ constant: this happens when the motion of the gas is along the magnetic field line. For instance, the initial turbulent motions of the gas can compress or de-compress the gas: when the gas is compressed (de-compressed) in the direction of the magnetic field lines, $B$ remains the same but the density increases (decreases). If the motion is perpendicular to $B$, the value of $B$ can increase or decrease for compression/decompression. However, this will just produce a constant scatter in the $B-\rho$ relationship around the mean if there is no preferred direction for the turbulence (isotropic turbulence). We expect the scatter around the mean to be small for smaller $\mu$, as the stronger magnetic tension/pressure suppresses compression/de-compression perpendicular to the B-filed direction. This is indeed observed in Figure~\ref{fig:brho1}.
 
This regime no longer exists at cores scales (high densities) when the gas motion is no longer isotropic, rather it is mostly compression due to the self-gravity of the cores \edit{under the influence of strong magnetic field}. Assuming that cores can be initially approximated as isothermal spheres embedded in a uniform magnetic field supported by thermal and turbulent pressure in the direction of the magnetic field lines, one can apply the derivation in \S~\ref{sec:b-rho} and Eq.~(\ref{eq:va}). After the initial phase of compression in the z-direction at constant $\Sigma$ and $B$, any further density increase is produced by compression perpendicular to the B-field lines, producing $B \propto \rho^{1/2}$ or $v_A={\rm const}(\rho)$. But what sets the constant value of $v_A \sim 0.5$ km/s observed across different scales and densities?

\begin{figure*}
  \def\thes{0.65}
  \centering 
\includegraphics[scale=\thes]{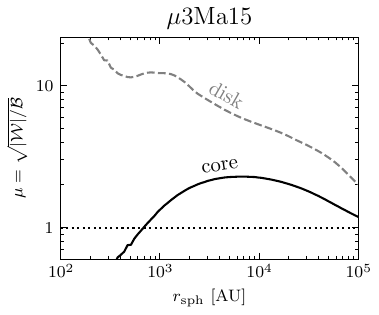}
\includegraphics[scale=\thes]{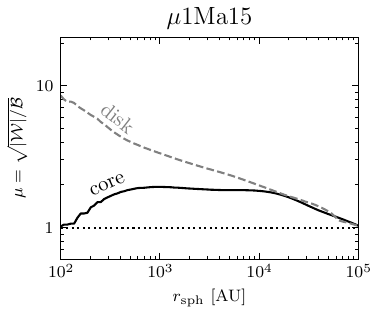}
\includegraphics[scale=\thes]{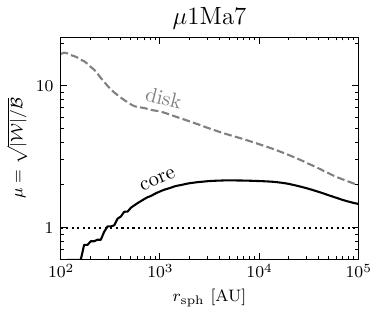}
\includegraphics[scale=\thes]{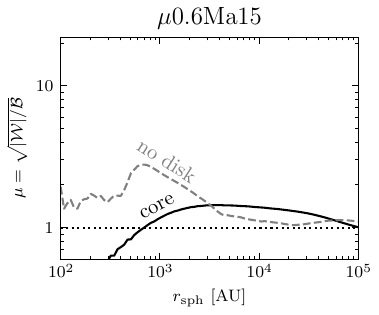}\\
\includegraphics[scale=\thes]{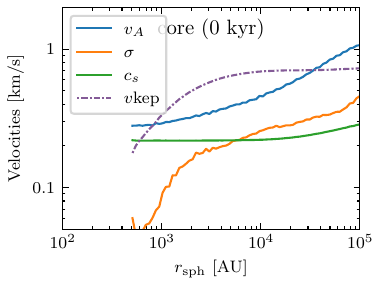}
\includegraphics[scale=\thes]{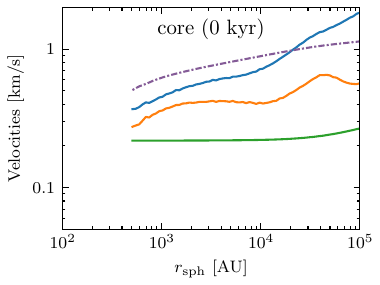}
\includegraphics[scale=\thes]{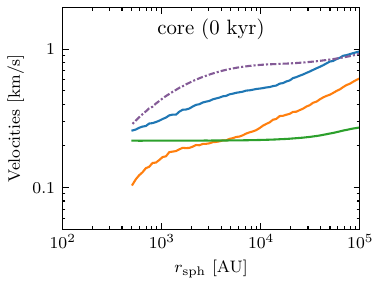}
\includegraphics[scale=\thes]{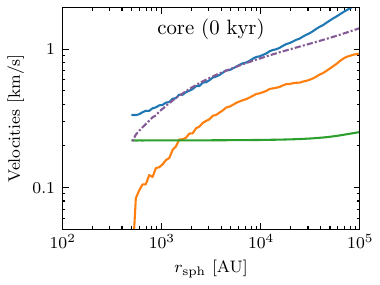}\\
\includegraphics[scale=\thes]{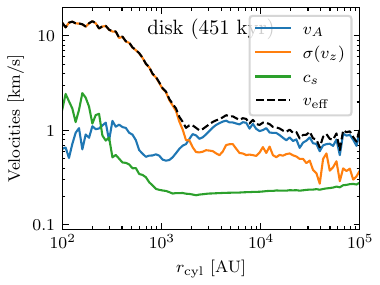}
\includegraphics[scale=\thes]{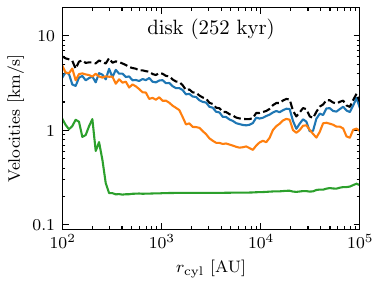}
\includegraphics[scale=\thes]{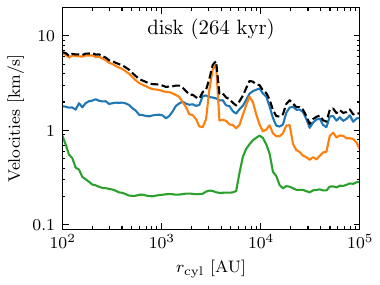}
\includegraphics[scale=\thes]{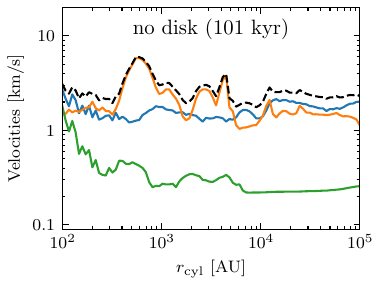}\\
\includegraphics[scale=\thes]{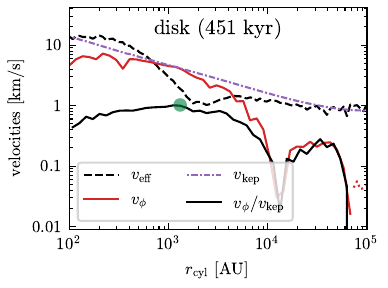}
\includegraphics[scale=\thes]{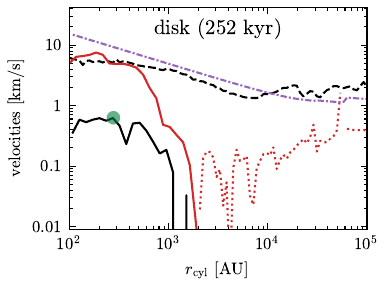}
\includegraphics[scale=\thes]{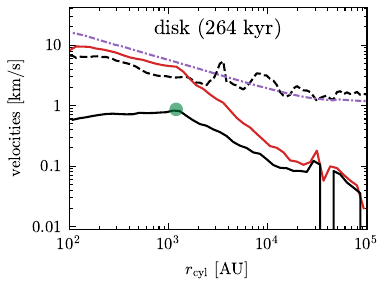}
\includegraphics[scale=\thes]{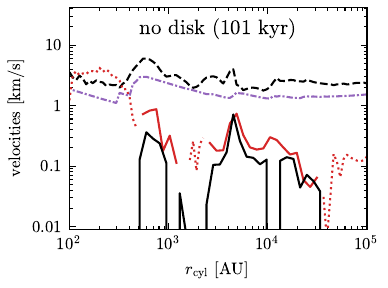}
\caption{\label{fig:vAvturb}
  The magnetic, turbulent, and thermal support of the cores and discs. From left to right are the cores \cHH{}, \cMH{}, \cML{}, and \cLH{}.
  Row 1: the mass-to-flux ratio radial profile of the gas at the core phase (initial time $t=0$) and disc phase (later stage) as a function of radius.
  Row 2: radial profiles of the \edit{mass-weighted} Alfven velocity, turbulence velocity, sound speed, and Keplerian velocity at the core phase as a function of the radius. 
  Row 3: radial profiles of the \edit{mass-weighted} Alfven velocity, $z$-component turbulent velocity, sound speed, and the effective velocity of support $v_{\rm eff} = (v_A^2 + \sigma(v_z)^2 + c_s^2)^{1/2}$ during the disc phase as a function of the radius in cylindrical coordinates.
  Row 4: radial profiles of the \edit{mass-weighted} effective velocity, the azimuthal velocity, the Keplerian velocity, \edit{and the Keplerianity $\beta_K = v_\phi / v_{\rm kep}$} during the disc phase as a function of the radius in cylindrical coordinates. \edit{The green dots at the last row mark the edge of the discs.} Quasi-Keplerian discs with $v_\phi \approx v_{\rm kep}$ extending to radii $r_{\rm cyl} \sim 300-5000$~AU form in all runs but \cLH{}.
  The cores are initially supported by magnetic pressure, while thermal and turbulent pressures are sub-dominant. The toroids that form in the cores are supported by both turbulent and magnetic pressure, with the former slightly dominating in the quasi-Keplerian discs found in the inner part of the cores ($\lesssim 1000$ AU) and the latter slightly dominating in the outer part (toroid). Thermal support is negligible in all the massive cores in this study.
  }
\end{figure*}

The value of $\sigma$ and $\mu$ in Eq.~(\ref{eq:va}) are not the initial values for the GMC, but rather the initial values for self-gravitating cores. If the cores are in quasi-hydrostatic equilibrium supported by turbulence and magnetic pressure, ${W} \sim ({\cal B} + {K}_{\rm turb})$. If the initial value of the magnetic pressure is comparable to or dominates over turbulence we expect $\mu \equiv \sqrt{|{W}/{\cal B}|} \sim 1$. Assuming that the core is marginally Jeans unstable and partially supported by turbulent pressure, the equivalent Jeans mass is given by
\begin{equation}\label{eq:mj}
  M_J = \frac{\pi^{5/2}}{6} \frac{\sigma^3}{(G^3 \rho)^{1/2}},
\end{equation}
where $\sigma$ is the rms of the turbulent velocity. In GMC simulations by \cite{He2019} the typical masses of prestellar cores (the most numerous cores in the core mass function) is $M \approx 1-5 \ \msun{}$, and $n = 10^7 \ \pcc{}$ is their typical gas number density. Using these values in Eq.~(\ref{eq:mj}) we get $\sigma \approx 0.6$~km/s. 
Note that the same value can be obtained by calculating the virial velocity of the core. 
Finally, using Eq.~(\ref{eq:va})  with $\mu \sim 1$ we have $v_A \approx 0.67 \sigma \approx 0.4~{\rm km/s}$, in agreement with the mean value for the whole GMC and for individual cores.
Because $\sigma{}$ is weakly dependent on $M_J$ and the core mass function in the GMC is dominated by small-mass cores with $M \sim 1 - 5 \ \msun{}$, most of the dense gas in the GMC is in cores with $\sigma \sim 0.6$~km/s. Therefore, in the high-density regime $v_A \approx 0.4$ km/s when averaged over the whole cloud.
Note, however, that our most massive cores have $M \sim 130 \msun{}$. Indeed, the B-$\rho$ phase diagram for the most massive core in our set ({\tt $\mu$3Ma15-large}) shows higher values of $v_A$, consistent with our interpretation. 
 
\subsection{Magnetic and Turbulent Support in the Cores and discs} 

The cores initially have nearly critical magnetic field strengths, with values of the mass-to-flux ratio radial profiles, $\mu(r)$, ranging from 0.5 to 2 from the inner to the outer part of the core (see the top row in Figure~\ref{fig:vAvturb}). The $\mu$ radial profiles in cores that form from GMCs with different initial magnetic field strengths are virtually indistinguishable from each other: the mass-to-flux ratio in the \cLH{} core is only slightly lower than that in the \cHH{} core. This is likely due to a selection effect because magnetically sub-critical ``clumps'' would fail to collapse and form a sink particle.
During the quasi-spherical initial collapse of the cores, the magnetic pressure \edit{($\propto \rho v_A^2$)} dominates over the turbulent \edit{($\propto \rho \sigma^2$)} and thermal pressure \edit{($\propto \rho c_s^2$)}. 
\edit{The fact that $v_{\rm kep} \sim v_A$ (second row of Figure~\ref{fig:vAvturb}) indicates the total magnetic energy in the core ($\sim M v_A^2/2$) is comparable with the total gravitational potential energy ($\sim 3G M^2 / (5R) \sim 3M v_{\rm kep}^2/5$). This suggests the core might be nearly in hydrostatic equilibrium with the magnetic pressure supporting the core against gravitational collapse.}

\edit{In the bottom row, we mark the edges of the discs with green dots, defined as the radius at which the Keplerianity parameter, $\beta_K = v_\phi / v_{\rm kep}$ begins to decrease from a constant value near unity. We also enforce the criteria that a disc must have $\beta_K \ge 0.5$ within its radius.}
The bottom two rows show that, as the core collapses, the turbulence is amplified especially in the inner parts of the disc/core. In the outer disc, turbulent kinetic energy and magnetic energy are nearly in equipartition. In the disc, the Alfven velocity ($v_A$) remains around 1~km/s, occasionally increasing to 5~km/s in the inner part. The effective velocity, $v_{\rm eff}^2 = v_A^2 + \sigma^2 + c_s^2$, always approaches the Keplerian velocity within the disc radius, while it drops significantly below $v_{\rm kep}$ outside of the disc radius. The disc remains quasi-Keplerian ($v_\phi \approx v_{\rm kep}$) up to a characteristic radius of about $1000 \pm 500$~AU in all the cores but the sub-critical one with $\mu =0.6$: in this run $v_\phi << v_{\rm kep}$ at all radii and a Keplerian disc fail to form (see Figure~\ref{fig:denprj}).

The grey dashed lines in the top row of Figure~\ref{fig:vAvturb} show the $\mu$ profiles of the discs at late times. The discs have $\mu(r) \sim 10$ in the inner parts where turbulent support is dominant, but $\mu(r)$ decreases to $\sim 2$ in the outer envelope. This decrease of $\mu(r)$ in the outer parts of cores has already been observed: \cite{Yen2022} has shown that the mass-to-flux ratio increases from 1-4 to 9-32 from 0.1~pc to 600~AU scales, which suggests that the magnetic field is partially decoupled from the neutral matter from large to small scales. The authors suggest non-ideal MHD (\eg, ambipolar diffusion) as the cause of this $\mu$ radial profile that allows the formation of a Keplerian disc. In our simulations, modelling of non-ideal MHD processes is not included in our equations, other than indirectly in our recipe for sink accretion: sinks accrete mass, momentum and angular momentum but not magnetic field. Hence, a deviation from flux-freezing is caused by the accumulation and subsequent diffusion of the magnetic field \citep{Zhao2011, Santos-Lima2012} within sink particles described before. The increase of $\mu$ in the inner part is mainly produced by the increase of the gravitational potential energy $|W|$ from the mass increase of the sink particle which, however, does not accumulate magnetic energy.

\subsection{Magnetic Braking Problem}
\label{sec:braking}
\begin{figure*}
  \centering
  \includegraphics[scale=0.62]{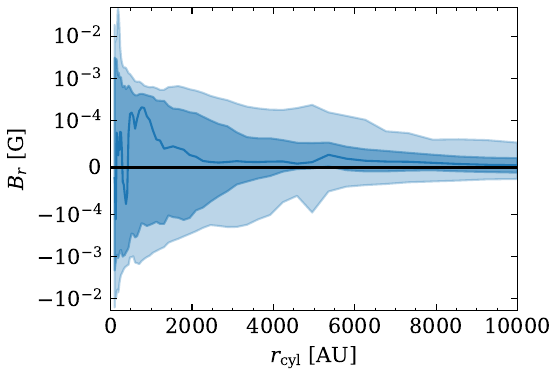}
  \includegraphics[scale=0.62]{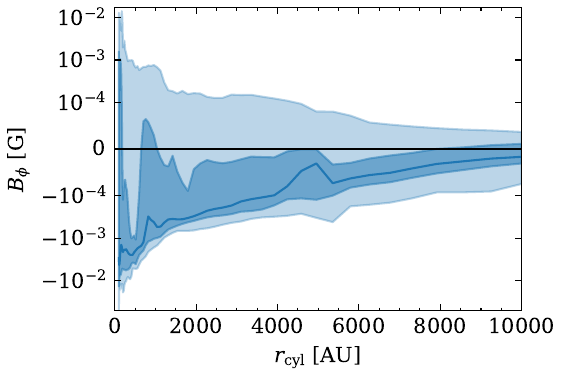}
  \includegraphics[scale=0.62]{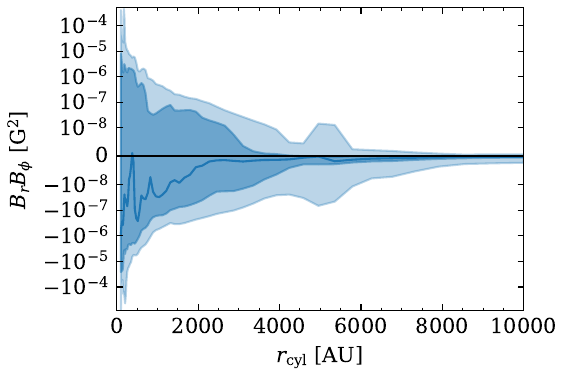}
  \caption{\label{fig:tbr-1}
  Explanation of the weak overall magnetic braking in disc \cHR{}. From left to right are the distributions of the radial component of the magnetic field, the azimuthal component of the magnetic field, and their product, which is proportional to $t_{\rm br}^{-1}$, as a function of disc radius in cylindrical coordination. 
  The thick curve displays the median values and the shaded area displays the 1-$\sigma$ and 3-$\sigma$ contours for the probability distribution function. %
  Positive and negative values of $B_r B_{\phi}$ (therefore $t_{\rm br}$) cancel each other, resulting in a small overall magnetic braking effect. }
\end{figure*}

As discussed before, the formation of a toroid or a disc can be suppressed or its radius reduced by magnetic braking. The spinning of the gas twists up the magnetic fields, creating a tension force that opposes rotation. The magnetic field exerts a Lorentz force per unit volume on the fluid element which, at a given radius $r$, can be written as
\begin{align}\label{eq:f}
  \mathbf{f} &= \frac{1}{4 \pi} \left[ (\nabla \times {\bf B}) \times {\bf B} \right] \\
			 &= \frac{1}{4 \pi r} \left[ {\bf B}_p \cdot \nabla_p (r B_{\phi}) \right] \hat{\phi} \\
			 &= \frac{1}{4 \pi} \left( B_r B_{\phi} + r B_r \frac{\partial}{\partial r} B_{\phi} + r B_z \frac{\partial}{\partial z} B_{\phi} \right) {\bf \hat\phi} \label{eq:f:3} \\
			 &\approx \frac{B_r B_{\phi}}{4 \pi} \hat{\mathbf{\phi}},
\end{align}
\edit{where the subscript $p$ denotes the poloidal component of the filed and $\nabla_p = (\frac{\partial}{\partial r} \hat{r}, \frac{\partial}{\partial \phi} \hat{\phi})$. }
We have only considered the $\phi$ component of the torque and assumed that the gradient of $B_{\phi}$ with respect to the $z$-direction vanishes due to symmetry. The second term in the parenthesis of Equation~(\ref{eq:f:3}) is negligible compared to the first term for a B-field with a azimuthal component that is nearly constant as a function or radius (\ie, $d\ln B_\phi/d\ln r < 1$). Therefore, we have
\begin{equation}
\frac{d}{d t} (\rho v_{\phi}) = - \frac{|{\bf f}|}{r} \approx -\frac{B_r B_{\phi}}{4 \pi r}.
\end{equation}
The magnetic braking time is defined as the characteristic timescale for the magnetic torque to remove completely the gas angular momentum:
\begin{equation}\label{eq:tbrake}
t_{\rm br} = \frac{\rho v_{\phi}}{-\frac{d}{d t}(\rho v_{\phi})} \approx \frac{4 \pi \rho v_{\phi} r}{B_r B_\phi}.
\end{equation}

To compare $t_{\rm br}$ with the dynamic timescale, we assume $B_r B_{\phi} \approx B^2$ and get
\begin{equation}
  t_{\rm br} \approx \frac{4 \pi \rho v_{\phi} r}{B^2} 
  = \left( \frac{v_\phi}{v_A} \right)^2 \frac{r}{v_\phi} 
  = \left( \frac{v_\phi}{v_A} \right)^2 t_{\rm cr}.\label{eq:tbr}
\end{equation}
This means that if a cloud has a magnetic field nearly in equipartition with gravitational potential energy and if the field is marginally wound up such that the poloidal and toroidal components become comparable, we expect \( t_{\rm br} \sim t_{\rm cr} \), \ie, the field is capable of stopping Keplerian rotation in a timescale of the order of the disc rotation period. For instance, at r = 500~AU in the disc of core \texttt{$\mu$3Ma15}, \(v_\phi \approx 6\)~km/s, \(v_A \approx 1\)~km/s, and \(t_{\rm cr} = 0.4\)~kyr. From Eq.~(\ref{eq:tbr}) we have \(t_{\rm br} \approx 36 t_{\rm cr} \approx 14\)~kyr that is significantly shorter of the disc lifetime $\gtrsim 300$~kyr. A timescale $t_{\rm br}$ a factor of two times longer is obtained if we assume $B_r B_\phi \approx B^2/2$.

\begin{figure*}
  \newcommand{\theh}{5in}
  \centering
  \includegraphics[]{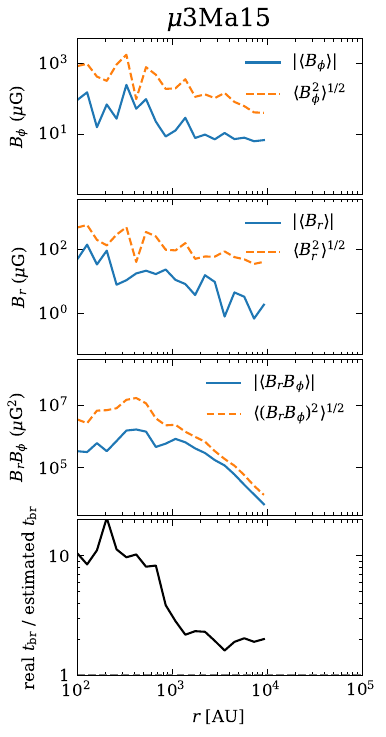}
  \includegraphics[trim={0.47in 0 0 0},clip]{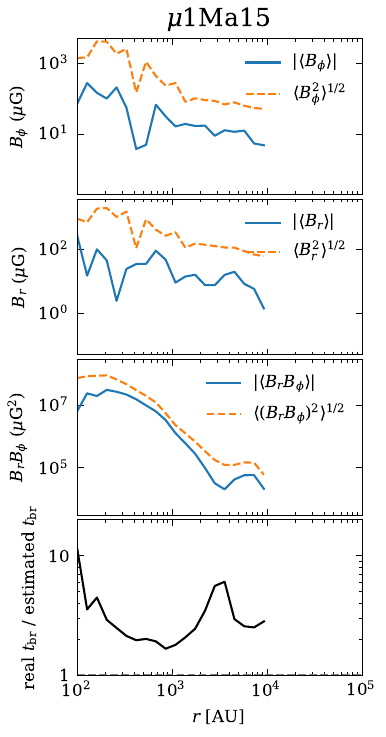}
  \includegraphics[trim={0.47in 0 0 0},clip]{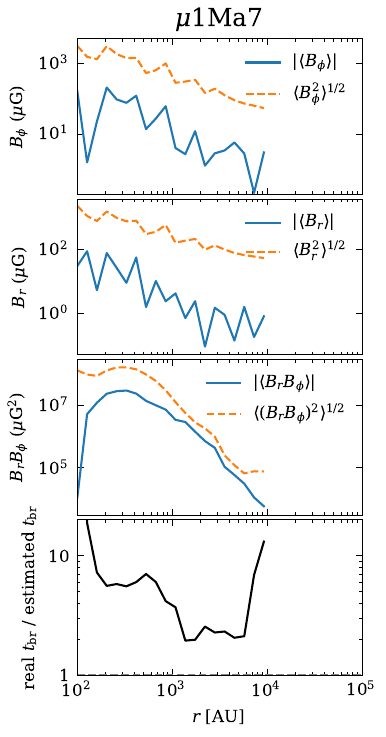}
  \caption{\label{fig:sigmaB} 
    The extremely turbulent and incoherent magnetic field on the circumstellar discs as a solution to the magnetic braking problem. The columns from left to right are discs \cHH{}, \cMH{}, and \cML{}, respectively.
    Top row: the mean and standard deviation of the azimuthal component of the B field as a function of the cylindrical radius.
    Second row: the mean and standard deviation of the radial component of the B field. 
    Third row: the mean and standard deviation of the products of the azimuthal and radial components of the B field. 
    Bottom row: The ratio of the magnetic braking time, $t_{\rm br} \propto | \langle B_r B_{\phi} \rangle |^{-1}$, to the naive estimation, proportional to $\langle (B_r B_\phi)^2 \rangle^{-1/2} \approx \langle B^{2} \rangle^{-1}$.
  } 
\end{figure*}

\edit{At least for the massive discs studied in this work, which are supported in the vertical direction by (supersonic) turbulent motions rather than thermal pressure, we argue that apparently contradictory results regarding the timescale for magnetic breaking and the critical $\mu$ value suppressing disc formation can be understood in terms of the turbulent and incoherent nature of magnetic fields in massive cores/toroids. } 
In Figure~\ref{fig:tbr-1}, we plot the distribution of the toroidal and poloidal components of the magnetic field, $B_{\phi}$ and $B_r$, \edit{in the cylindrical coordinate at a given distance to the disc centre} for core {\tt $\mu$3Ma7hires}. \edit{The coordinate system is the same as that defined in the previous section. At a given radius in the cylindrical frame, we pick all the cells in a concentric and superposed ring and compute each component of the $B$ field weighted by gas mass. The median, $1-\sigma$, and $3-\sigma$ contours for the distribution of the radial component, $B_r$, the azimuthal component, $B_\phi$, and their product are then plotted in the three panels.} 
We can see that while the azimuthal component of the magnetic field is mostly directional, the radial component, on the other hand, evenly scatters around zero. The radial component $B_r$ is extremely turbulent.
\edit{and the turbulent root-mean-square (rms), $<B_r^2>^{1/2}$, is roughly 10x higher than the mean (see second row of Figure~\ref{fig:sigmaB}). }
Consequently, the magnetic torque, proportional to $B_r B_{\phi}$, also scatters around zero. This results in the torque exerted on the gas cancelling itself out, greatly weakening the magnetic breaking effect. This reduction of roughly a factor of ten of the torque increases the breaking timescale $t_{\rm bf}$ by the same factor, ultimately 
influencing the longevity and stability of the discs \edit{compared to the case where the magnetic field lines are perfectly coherent and aligned with the spin axis of the disc/core}.

As demonstrated by Figure~\ref{fig:sigmaB}, showing the radial profile of the \edit{volume-weighted} mean and rms of $B_{\rm \phi}$, $B_{\rm r}$ and $B_{\rm r}B_{\phi}$ for three cores in Table~\ref{tab:1}, this incoherent character of the magnetic field does not depend on the initial turbulence of the GMC, which ranges from ${\cal M} = 7$ to $15$, nor on the magnetic strength, which ranges from marginally-supercritical to critical ($1<\mu <3$).
Disc \cMH{} has the strongest magnetic strength (lowest $\mu$) in the disc and also the smallest degree of turbulence-induced weakening of the magnetic breaking effect among the three discs inspected.
We have also performed a resolution study, presented in Appendix~\ref{sec:reso}, to rule out resolution effects as a cause for the formation of large discs.

\section{Discussion}\label{sec:disc}

\edit{The exponent of $1/2$ we find for the $B - \rho$ power-law relationship in the range of densities between $10^5$~cm$^{3}$ and $10^9$~cm$^{3}$ indicates that the cloud collapse at these relatively large scales (compared to protostellar core scales, \ie{}, $> 10^4$~ AU) have cylindrical/filamentary geometry rather than spherical. This is true across all GMCs in our sample, irrespective of whether the turbulent energy is the dominant support against gravity (${\cal M}_A = 5$ case) or the magnetic energy dominates (${\cal M}_A = 1$ case). 
Our findings seem in contrast with the conclusion of a previous study by \cite{Mocz2017}, which proposed that the collapse is isotropic ($B \propto \rho^{2/3}$) in environments when turbulence dominates over the magnetic field and transitions into anisotropic when the magnetic energy dominates. We identify two main factors contributing to this discrepancy. First, the range of densities investigated differs; our analysis spans up to $10^9 \ \pcc{}$, whereas Mocz et al. reached densities around $10^{11} \ \pcc{}$. Second, the authors point out that the $B-\rho$ relation, when measured on a cell-by-cell basis -- a method more akin to ours -- exhibits a flatter power-law slope.
}

\subsection{Previous ideal-MHD calculations of massive core collapse}

\edit{ A fundamental difference between hydrodynamic and magneto-hydrodynamic prestellar cores comes from the evolution of angular momentum. In the absence of a significant magnetic field, the angular momentum is essentially conserved and becomes a dominant support againt gravitational collapse, drastically affecting the evolution of the cores. In a magnetised core, however, the situation is different; angular momentum can be exchanged between fluid particles because of magnetic force/tension. In the aligned configuration between the magnetic field and the cloud rotation axis, the magnetic braking time can become so short that the formation of the centrifugally supported discs can be even entirely prevented. }

\edit{ Recent numerical studies have explored how magnetic configurations/geometries, \ie, misalignment, may reduce the power of magnetic braking. \cite{Hennebelle2009} report the first discovery that centrifugal discs more easily form in cores in which the magnetic field and the angular momentum vector are misaligned. \cite{Joos2012}, \cite{Li2013c}, and \cite{Hirano2020} report more quantitative results that are consistent with \cite{Hennebelle2009}. \cite{Joos2012} show that, in the case where the magnetic field and angular momentum are perpendicular to each other, the mean specific angular momentum of the central region of a core is about two times larger than that in the parallel case.
\cite{Tsukamoto2018}, however, claim that in the prestellar collapse phase, the presence of misalignment seems to enhance the angular momentum removal from the central region due to stronger magnetic braking, contradicting before-mentioned works. Similarly, \cite{Li2013c} argue that misalignment alone does not enable the majority of disc formation. Future observations should be able to provide more evidence to disentangle this apparent discrepancy between simulation results. }

\edit{Several studies have investigated the effects of turbulent initial magnetic fields on disc formation \citep[e.g.][]{Joos2013, Li2014, Fielding2015, Gray2018, Lewis2018, Lam2019}. \cite{Joos2013} explored the impact of turbulence with high resolution and suggested that the impact of turbulence is limited. Nonetheless, \cite{Tsukamoto2023} argue that turbulent diffusion, either as a numeric effect or non-ideal MHD effect, may not play a significant role in circumstellar disc size evolution. }

\edit{ The calculations we present here extend the investigation of star formation under ideal MHD in two major ways beyond the studies mentioned above. Unlike in most of the above MHD simulations, which typically track the collapse of clouds around $100 \msun{}$, our work zoom-in at high resolution (tens of AU) a few $30-100$~M$_\odot$ cores within the realistic environment of a much larger GMC. Specifically, we model the formation of massive prestellar cores, with masses between tens to one hundred solar masses, emerging from the collapse of GMCs with masses $\sim 10^4 \msun{}$ and different level of magnetisation and turbulent energy. Despite of the computational challenges posed by simulating such large masses, we achieve a relatively high spatial resolution (between 10 and 60 AU), sufficient to resolve discs with radii $\gtrsim 200$ AU, and densities $\gtrsim 10^{9} \pcc{}$. The adoption of realistic initial and boundary conditions accurately captures the thermal, turbulent, and magnetic energy configurations essential for comparing their relative roles in the star formation process. Because our setup differs in mass scale and boundary/initial condition from most of previous studies we find results that at times seem to contradict previous results established in he literature. For instance, we find that large ($R \sim 300 - 1000$ AU), rotationally supported discs can indeed form from the ideal-MHD collapse of prestellar cores, even under conditions where the cores are magnetically near-critical or critical. We motivate this results as due to the highly turbulent nature of the magnetic field within the discs, that are supported in the vertical direction by turbulent motions rather than thermal pressure. This is also contrary to previous findings for smaller mass discs that instead are supported by thermal pressure.} 

\subsection{Previous non-ideal MHD simulations of massive core collapse }

\edit{Non-ideal MHD effects may play a crucial role in shaping the evolution of discs, impacting their size through the decoupling of magnetic fields and gas, thereby reducing the magnetic braking efficiency within the disc \citep[see][for a review]{Tsukamoto2023}. First of all, relying solely on Ohmic dissipation proves insufficient in driving the formation of large circumstellar discs during the Class 0 phase (the first protostellar objects observed post-collapse within prestellar cores) \citep{Machida2011}. This limitation arises from the fact that to significantly diffuse magnetic fields, Ohmic heating would elevate the core's inner region's temperature to a level where thermal energy matches gravitational energy, hindering the collapse of the core as a whole.}

\edit{Ambipolar diffusion is a more effective magnetic diffusion process that weakens the effects of magnetic field in the disc and envelope of a protostar by allowing the ions to slip past the neutrals. 3D simulations have shown that ambipolar diffusion allows the formation of a relatively massive disc with a size of tens of AU in the early disc formation phase, despite a relatively strong magnetic field \citep[e.g.][]{Masson2016,Zhao2018,Tsukamoto2020}. 
By comparing simulations for a coherently rotating core and a turbulent core, \cite{Santos-Lima2012} suggested that turbulence causes small-scale magnetic reconnection and provides an effective mechanism for magnetic diffusion that can remove magnetic flux out of the disc progenitor at timescales comparable to the collapse, allowing the formation of discs with sizes $\sim 100$ AU. Similar results are reported by \cite{Seifried2013}. 
\cite{Mignon-Risse2021} further suggest that, under moderate magnetic fields, massive protostellar discs with radius $\gtrsim 100$ AU can only be reproduced in the presence of ambipolar diffusion, even in a turbulent medium. Similar results are reported by \cite{Kolligan2018, Commercon2022, Mignon-Risse2023}, some reporting discs with radius even $\gtrsim 500$ AU. 
Contrastingly, \cite{Wurster2016a} argue that discs do not form in simulations in the presence of strong magnetic fields even when ambipolar diffusion and Ohmic resistivity are included. They argue that large disc formation occurs only under weaker, anti-aligned fields influenced by the Hall effect. 
Additionally, the role of cosmic-ray complicates this scenario, as high rates of ionization, shown by  \cite{Kuffmeier2020,Zhao2016}, can suppress the formation of discs. \cite{Wurster2019} suggests that non-ideal MHD effects moderate the magnetic field in the circumstellar environment, leading to a more constrained range of disc field strengths compared to the ideal-MHD case, which facilitates disc formation. After all, future observations aimed at detecting ion-neutral drift in prestellar cores will be essential in quantifying ambipolar diffusion's significance in disc formation processes.
}

\edit{While our simulations operate within the ideal MHD framework and do not encompass non-ideal MHD effects, the insights garnered in this work on the collapse of massive cores within realistic GMCs is novel and it is a first step toward understanding the interplay between non-ideal MHD phenomena and the role of incoherent, diffused magnetic field lines on the star formation processes. This omission, due to the scope of our investigation, does not diminish the relevance of our findings, but rather it sets the stage for future studies to delve into the comparative impacts of non-ideal vs ideal MHD processes.}

\section{Summary}
\label{sec:sum}

We have studied the collapse of strongly magnetised prestellar cores from a set of zoom-in radiation-MHD simulations of star formation in GMCs. The study includes a suite of six simulations of prestellar cores in molecular clouds with varying magnetization ($\mu_0 = 3, 1, 0.6$) and turbulence (${\cal M} = 15, 7$). We come to the following main conclusions:

\begin{enumerate}[wide,labelwidth=!,itemindent=!,labelindent=0pt,leftmargin=0em,label=\arabic*.]
\item
  We find a universal $B$-$\rho$ relationship, $B \approx 86~\mu{\rm G} \ (n/10^5~{\rm cm}^{-3})^\frac{1}{2}$, implying a constant Alfven velocity $v_A \approx 0.5 km/s$, for number density in the range between $10^5 \ {\rm cm}^{-3}$ and $10^9 \ {\rm cm}^{-3}$ in the evolution of magnetically critical or super-critical GMCs, regardless of the initial magnetic intensity or cloud size (see Figure~\ref{fig:brho1}).
  This value of $v_A$ roughly equals the virial velocity of a core with a mass at the peak of the core mass function.
\item \edit{On the large scales of collapsed cores ($r > 10^4$ AU), the scaling of the magnetic field with gas density (looking at it on a simulation cell-by-cell basis) is close to $B \propto n^{1/2}$. This is found at densities between $10^5 \ \pcc{}$ and $10^9 \ \pcc{}$ across all GMCs, irrespective of whether the turbulent energy dominates (${\cal M}_A = 5$) or is comparable to the magnetic energy (\textbf{${\cal M}_A = 1$)}}. 
\item 
  Keplerian circumstellar discs can form in critical and supercritical cores. Subcritical cores, however, fragment into numerous low-mass clumps that undergo direct collapse without any accretion, causing the absence of circumstellar discs (Figure~\ref{fig:denprj}).
\item
  Large discs can form in magnetically (near-)critical cores because the magnetic field is extremely turbulent and incoherent, which reduces the effect of magnetic braking by roughly one order of magnitude (see Figure~\ref{fig:sigmaB}) \edit{compared to the scenario where the magnetic field is coherent and perfectly aligned with the rotational axis of the core}. The turbulent magnetic field is caused by the supersonic turbulent motion of the gas in the disc, which is a result of the non-axisymmetric accretion of the gas enhanced by gravitational collapse.
\item
The cores at the initial phase are near critical with $\mu$ ranging between 0.6 and 2 (Figure~\ref{fig:vAvturb}). These cores are initially predominantly supported by magnetic pressure, with thermal and turbulent pressures playing lesser roles. The discs that form in the cores are supported by both turbulent and magnetic pressure, with the former slightly dominating in the inner part ($\lesssim 1000$ AU) and the latter slightly dominating in the outer part. \edit{This result, which is contrary to what is typically found in the literature for smaller mass discs, was also found in Paper~I and is due to the large masses of the cores studied in this work (30-300 M$_\odot$) and in part to the realistic boundary conditions that generate a continuous source of turbulence through asymmetric inflow.} In all our massive discs thermal pressure support is negligible.
\end{enumerate}

Despite the success of our model in reproducing some key features of circumstellar disc formation, we acknowledge that our study has known limitations. For instance, we do not account for the effects of protostellar outflow/jet feedback, radiation pressure, and radiative heating by low-energy photons on disc evolution. 
Feedback mechanisms could significantly alter the disc structure and dynamics. \edit{And finally, perhaps the most dramatic limitation at these small scales is the ideal-MHD approximation.} In future work, we plan to incorporate more realistic stellar feedback prescriptions in our code and possibly non ideal-MHD effects to overcome these limitations and improve the accuracy of our circumstellar disc models.

\section*{Acknowledgements}

CCH acknowledges the support by the NASA FINESST grant 80NSSC21K1850. The authors acknowledge the support of the NASA grant 80NSSC18K0527. The authors also acknowledge the University of Maryland supercomputing resources (http://hpcc.umd.edu) made available for conducting the research reported in this paper.

\section*{Data Availability}

The data underlying this article were accessed from the University of Maryland supercomputing resources (\url{http://hpcc.umd.edu}). The derived data generated in this research will be shared upon reasonable request to the corresponding authors. The software used to do the analysis in this paper is {\sc ramtools}, a toolkit to post-process \ramses{} simulations and is based on the {\sc yt} toolkit (\url{https://yt-project.org/doc/index.html}), available to download from \url{https://chongchonghe.github.io/ramtools-pages/}.

\bibliographystyle{mnras}
\bibliography{BIB_HE,disk} %

\appendix

\section{Mass-to-flux ratio of a non-singular isothermal sphere}
\label{sec:mu}

The density profile of a non-singular isothermal core in hydrostatic equilibrium is given by
\begin{equation}\label{eq:iso}
  \rho(r) = \frac{\rho_0}{1+(\frac{r}{r_c})^2}.
\end{equation}
Defining the dimensionless radius $\xi \equiv r/r_c$, we can show that the mass of the gas within $\xi$ is $M(\xi)=4\pi \rho_0 r_c^3 (\xi -\arctan \xi)$. 
Assuming a parallel magnetic field threading the midplane of the sphere with the magnetic strength proportional to $\rho^{1/2}$ and equal to $B_0$ a the centre, then we have
\begin{equation}\label{eq:A:Br}
  \Phi_B = \int_0^{\xi_1} 2 \pi r_c \xi \frac{B_0}{\sqrt{1 + \xi^2}} r_c d \xi = 2 \pi B_0 r_c^2 \left( \sqrt{1+\xi_1^2}-1 \right). 
\end{equation}
The magnetic critical mass is given by Equation~(\ref{eq:mphi}). 

The gravitational binding energy of a core with radius $\xi_1$ is given by this integral
\begin{align}
  W &= -\int_0^{\xi_1} G M(\xi) 4 \pi r \rho(r) r_c d\xi \nonumber \\
  &= -(4\pi)^2 G \rho_0^2 r_c^5 \int_0^{\xi_1} \frac{\xi(\xi - \arctan \xi)}{1+\xi^2} d\xi.
\end{align}
The total magnetic energy inside radius $\xi_1$ is 
\begin{align}\label{eq:A:eB}
  e_B &= \int_0^{\xi_1} 2 \pi r_c^2 \xi d\xi 2r_c\sqrt{\xi_1^2 - \xi^2} \frac{1}{8\pi} \frac{B_0^2}{1+\xi^2} \nonumber \\
&= \frac{1}{2}B_0^2 r_c^3 \int_0^{\xi_1} \frac{\xi \sqrt{\xi_1^2 + \xi^2}}{1+\xi^2} d\xi.
\end{align}

\begin{figure}
  \centering
  \includegraphics[]{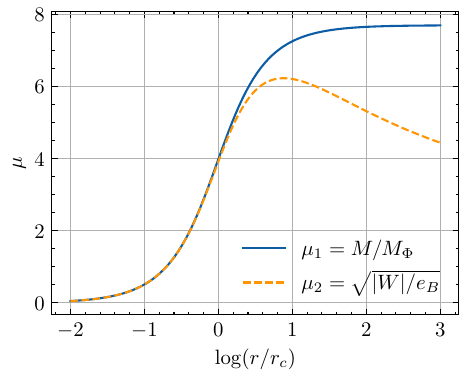}\\
  \vspace{-0.6cm}
  \includegraphics[]{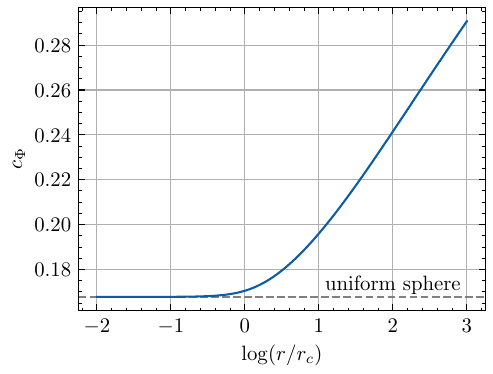}
  \caption{\label{fig:A:mu12} Comparing two definitions of the relative importance of the gravitational and magnetic forces in a non-singular isothermal sphere as a function of its radius: the relative mass-to-flux ratio $\mu_1 = M/M_{\Phi}$ and the square root of the binding to magnetic energy $\mu_2 = \sqrt{|W|/e_B}$. In the former case, the geometrical factor $c_\Phi$ in Eq.~(\ref{eq:mphi}) equals $1/\sqrt{2}$. The value of $c_\Phi$ required for the two definitions of $\mu$ to be equivalent to each other (\ie, $\mu_1 = \mu_2$) is shown in the bottom panel, showing that $c_\Phi$ increases as the core becomes more centrally concentrated.
}
\end{figure}

We plot $\mu_1 = M/M_{\Phi}$ and $\mu_2 = \sqrt{|W|/e_B}$ for an isothermal core with $\rho_{0} = 10^9 m_p~{\rm cm}^{-3}$, $r_c = 1000~{\rm AU}$, and $B_0 = 0.01$ Gauss in the top panel of Figure~\ref{fig:A:mu12}. In the bottom panel, we show the value of the equivalent geometrical factor $c_\Phi=\sqrt{G}M/(\Phi_B \mu_2)$ in Equation~(\ref{eq:mphi}), required to have $\mu_2 = \mu_1$.

\section{Resolution study}
\label{sec:reso}

\begin{figure*}
  \centering
  \includegraphics[width=0.8\columnwidth]{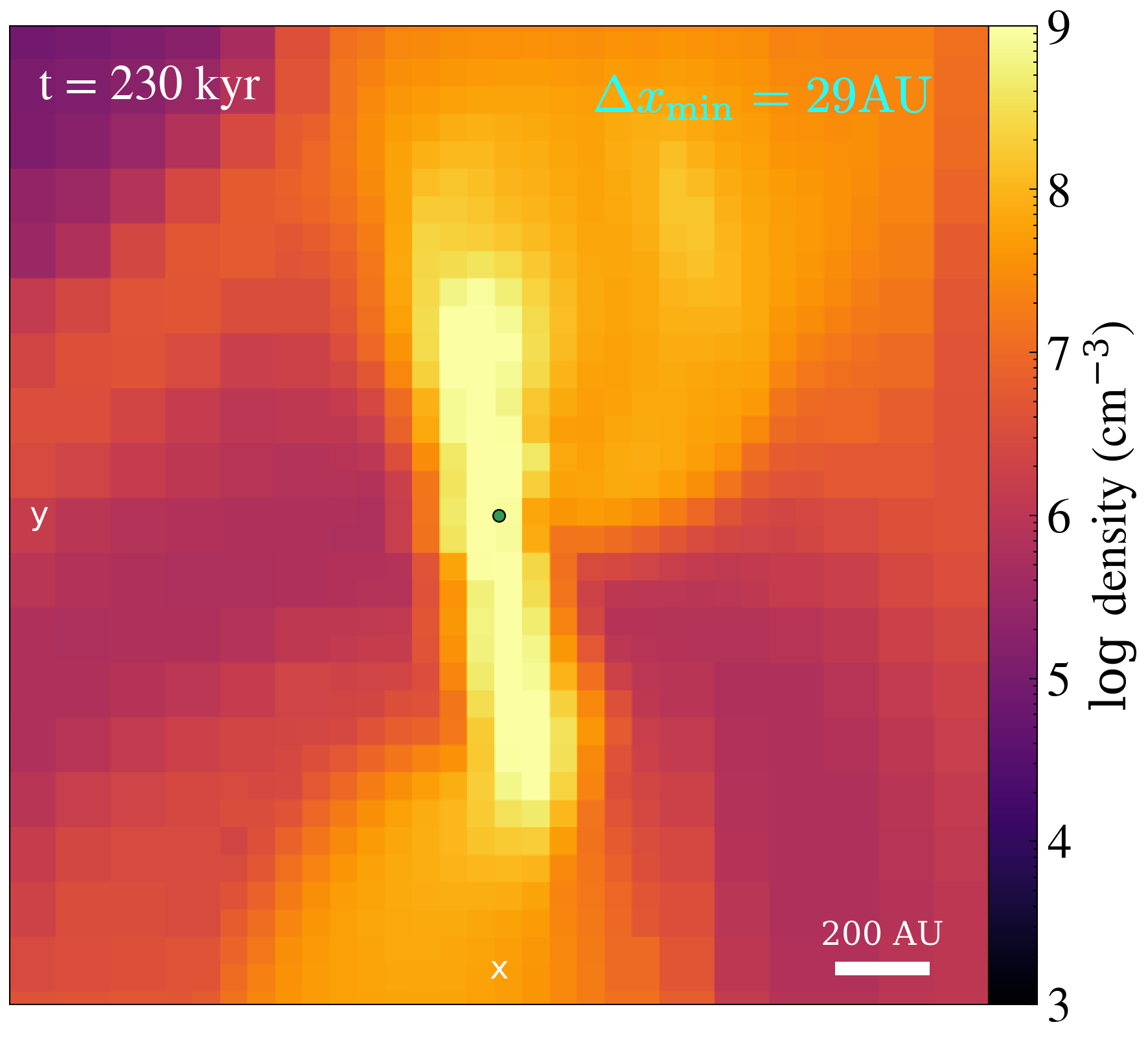}
  \includegraphics[width=0.8\columnwidth]{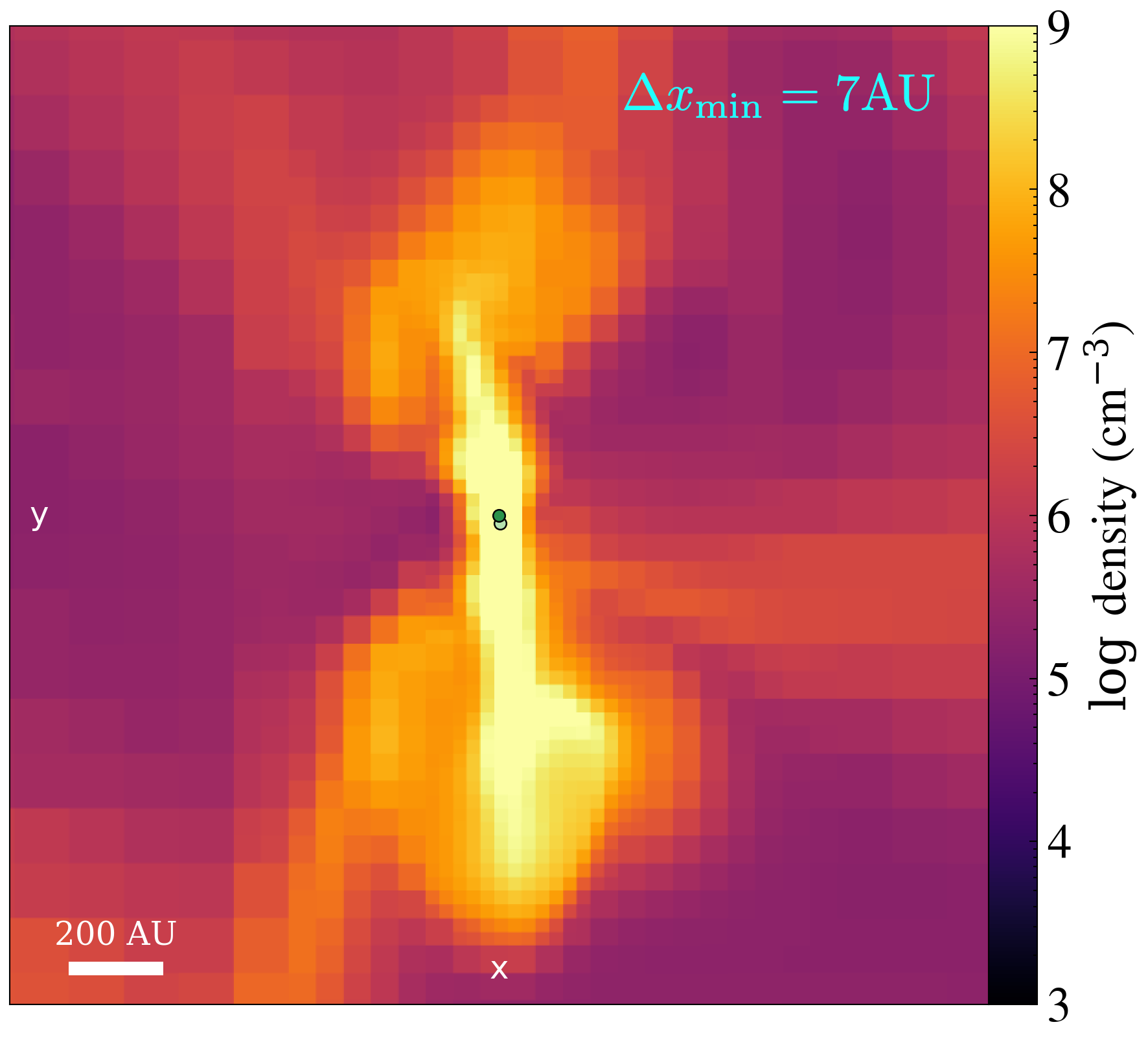}
  \caption{\label{fig:reso}
  A comparison of disc structure between two simulations with different resolutions. The figure shows an edge-on view of the discs from two simulations with the same initial conditions but varying resolutions. The left panel is run {\tt $\mu$3M7} with $l = 18$ and $\Delta x_{\rm min} = 29 {\rm AU}$, and the right panel is run {\tt $\mu$3M7-hires} with $l=20$ and $\Delta x_{\rm min} = 7 {\rm AU}$. The disc structure and size are similar in both runs, indicating numerical convergence of the resolution \edit{on the resolution of these large discs}.
  }
\end{figure*}

In this section, we explore the numerical convergence of the results by comparing {\tt $\mu$3M7} with {\tt $\mu$3Ma7-hires} which differs only in resolution, with the latter being 4 times higher in linear resolution. \edit{The resolutions are 29 AU for {\tt $\mu$3M7} and 7 AU for {\tt $\mu$3M7-hires}, and the corresponding sink formation densities are $3.6 \times 10^9 \ {\rm cm^{-3}}$ and $5.7 \times 10^{10} \ {\rm cm^{-3}}$.} We show edge-on projections of the discs from these two simulations in Figure~\ref{fig:reso}. 
\edit{With a resolution four times higher, the disc radius remains approximately unchanged at $\sim 600$ AU, and the disc's thickness decreases by less than 30 per cent. Our holistic approach in capturing realistic initial and boundary conditions constrains us from achieving a higher resolution. However, this resolution is sufficient to resolve these massive, large discs. Despite limited statistics, we can tentatively exclude resolution effects as a significant factor in the formation of these large discs.} 

\bsp	%
\label{lastpage}
\end{document}